\documentclass[
onecolumn,
nofootinbib,
amsmath,amssymb,amsthm,
preprint,
tightenlines
]{revtex4-2}

\usepackage[utf8]{inputenc}
\usepackage{mathtools}
\usepackage{amsmath}
\usepackage{amsthm}
\usepackage{amsbsy}
\usepackage{amssymb}
\usepackage{amscd}
\usepackage{amsfonts}
\usepackage{bm}
\usepackage{graphics}
\usepackage{graphicx}
\usepackage[plain]{fancyref}
\usepackage{soul}
\usepackage{hyperref}
\setcitestyle{numbers,sort&compress}
\usepackage{verbatim}
\usepackage[margin=25mm]{geometry}
\usepackage{paralist}
\usepackage{wasysym}
\usepackage{lineno}

\graphicspath{ {figs/} }

\newcommand*{\fancyrefapplabelprefix}{app}
\frefformat{plain}{\fancyrefapplabelprefix}{appendix~#1}
\Frefformat{plain}{\fancyrefapplabelprefix}{Appendix~#1}
\frefformat{main}{\fancyrefapplabelprefix}{appendix~#1}
\Frefformat{main}{\fancyrefapplabelprefix}{Appendix~#1}

\newcommand{\dihedral}[1]{\mathcal{D}_{#1}}

\newcommand{\reflOp}{\bm{\sigma}}
\newcommand{\refl}[1]{\reflOp_{#1}}

\newcommand{\SOThree}{SO\left(3\right)}

\newcommand{\OThree}{O\left(3\right)}
\newcommand{\bvec}{\hat{\mathbf{e}}}
\newcommand{\eone}{\bvec_1}
\newcommand{\etwo}{\bvec_2}
\newcommand{\ethree}{\bvec_3}
\newcommand{\rotMatrix}{\mathbf{Q}}
\newcommand{\rotAbout}[1]{\rotMatrix_{#1}}
\newcommand{\rotate}[2]{\rotAbout{#1}\left(#2\right)}
\newcommand{\skwMatrix}{\mathbf{A}}
\newcommand{\skwAbout}[1]{\skwMatrix_{#1}}
\newcommand{\unitVec}{\hat{\mathbf{u}}}
\newcommand{\otherUnitVec}{\hat{\mathbf{v}}}
\newcommand{\subgroup}{\leq}
\newcommand{\auxuj}[1]{\hat{u}_{#1}}
\DeclareMathOperator{\Orb}{Orb}

\newcommand{\orbit}[2]{\Orb_{#2} #1}
\newcommand{\group}{\mathcal{G}}
\newcommand{\otherGroup}{\mathcal{H}}
\newcommand{\orthoTrans}{\mathbf{O}}
\newcommand{\genBy}[1]{\left\langle #1 \right\rangle}
\newcommand{\lpt}{\mathbf{q}}
\newcommand{\tpt}{\mathbf{p}}
\newcommand{\NCreases}{N}
\newcommand{\creaseVec}{\hat{\mathbf{c}}}
\newcommand{\CreaseVec}{\hat{\mathbf{b}}}
\newcommand{\foldAngle}{\phi}
\newcommand{\partFoldAngle}{\rho}

\newcommand{\defMap}{\bm{\Phi}}

\newcommand{\F}{\mathbf{F}}
\newcommand{\facet}[1]{\cBody^{\left(#1\right)}}
\newcommand{\Facet}[1]{\rBody^{\left(#1\right)}}
\newcommand{\sectAngle}{\alpha}
\newcommand{\convToFoldAngle}{f_{\foldAngle}}

\newcommand{\elevAngle}{\upsilon}

\newcommand{\xPMag}{x}
\newcommand{\xP}{\mathbf{\xPMag}}
\newcommand{\xCMag}{y}
\newcommand{\xC}{\mathbf{\xCMag}}

\newcommand{\unNormalVec}{\mathbf{n}}

\newcommand{\heaviside}[1]{H\left(#1\right)}
\DeclareMathOperator{\Grad}{Grad}
\newcommand{\cBody}{\Omega}

\newcommand{\rBody}{\cBody_0}

\newcommand{\Reals}{\mathbb{R}}

\newcommand{\Integers}{\mathbb{Z}}
\newcommand{\Naturals}{\mathbb{N}}

\newcommand{\genAngle}{\varphi}
\newcommand{\Iden}{\mathbf{I}}
\newcommand{\outProd}[2]{#1 \otimes #2}
\newcommand{\set}[1]{\left\{ #1 \right\}}
\newcommand{\generic}{\Box}
\DeclareMathOperator{\sgn}{sgn}
\DeclareMathOperator{\cSin}{sin}
\DeclareMathOperator{\cCos}{cos}
\DeclareMathOperator{\cTan}{tan}
\newcommand{\numer}{\mathcal{Y}}
\newcommand{\denom}{\mathcal{X}}
\newcommand{\argSign}{\mathcal{S}}

\newcommand{\sVol}{\mathcal{V}}
\makeatletter
\newcommand*{\gnuplotinput}[2][1.0]{%
	\begingroup
	\let\@gnplt@input@includegraphics=\includegraphics
	\def\includegraphics##1{\@gnplt@input@includegraphics[scale=#1]{#2}}%
	\let\@gnplt@input@picture=\picture
	\def\picture{\unitlength=#1\unitlength\relax\@gnplt@input@picture}%
	\input{#2}%
	\endgroup
}
\makeatother

\theoremstyle{remark}

\newcommand*{\fancyrefproplabelprefix}{prop}
\frefformat{plain}{\fancyrefproplabelprefix}{proposition~#1}
\Frefformat{plain}{\fancyrefproplabelprefix}{Proposition~#1}
\frefformat{main}{\fancyrefproplabelprefix}{proposition~#1}
\Frefformat{main}{\fancyrefproplabelprefix}{Proposition~#1}

\newcommand*{\fancyreflemlabelprefix}{lem}
\frefformat{plain}{\fancyreflemlabelprefix}{lemma~#1}
\Frefformat{plain}{\fancyreflemlabelprefix}{Lemma~#1}
\frefformat{main}{\fancyreflemlabelprefix}{lemma~#1}
\Frefformat{main}{\fancyreflemlabelprefix}{Lemma~#1}

\newenvironment{hlbreakable}%
{}%
{}

\begin{document}

\title{Lagrangian approach to origami vertex analysis: Kinematics}
\author{Matthew Grasinger}
\email{matthew.grasinger.1@us.af.mil}
\affiliation{Materials \& Manufacturing Directorate, Air Force Research Laboratory}

\author{Andrew Gillman}
\affiliation{Materials \& Manufacturing Directorate, Air Force Research Laboratory}

\author{Philip Buskohl}
\email{philip.buskohl.1@us.af.mil}
\affiliation{Materials \& Manufacturing Directorate, Air Force Research Laboratory}

\preprint{To appear in Philosophical Transactions of the Royal Society A, doi: \href{https://doi.org/10.1098/rsta.2024.0203}{10.1098/rsta.2024.0203}.}

\begin{abstract}
The use of origami in engineering has significantly expanded in recent years, spanning deployable structures across scales, folding robotics, and mechanical metamaterials. However, finding foldable paths can be a formidable task as the kinematics are determined by a nonlinear system of equations, often with several degrees of freedom. In this work, we leverage a Lagrangian approach to derive reduced-order compatibility conditions for rigid-facet origami vertices with reflection and rotational symmetries. Then, using the reduced-order conditions, we derive exact, multi-degree of freedom solutions for degree 6 and degree 8 vertices with prescribed symmetries. The exact kinematic solutions allow us to efficiently investigate the topology of allowable kinematics, including the consideration of a self-contact constraint, and then visually interpret the role of geometric design parameters on these admissible fold paths by monitoring the change in the kinematic topology. We then introduce a procedure to construct lower symmetry kinematic solutions by breaking symmetry of higher order kinematic solutions in a systematic way that preserves compatibility. The multi-degree of freedom solutions discovered here should assist with building intuition of the kinematic feasibility of higher degree origami vertices and also facilitate the development of new algorithmic procedures for origami-engineering design.
\end{abstract}

\maketitle

\section{Introduction}
\label{sec:intro}
The use of origami principles in engineering has grown rapidly, highlighting its broad utility, versatility, and elegance.
The scale invariance of these principles has led to the adoption of origami design at length scales as large as deployable space structures~\cite{schenk2014review,zirbel2013accommodating} and shelters~\cite{verzoni2022transition,melancon2021multistable}, to smaller-scale biomedical applications~\cite{kuribayashi2006self,andersen2009self,velvaluri2021origami,jiang2019rationally}, nanorobotics~\cite{liu2023light}, and nanoscale devices~\cite{cho2011nanoscale,grasingerIPnanoscale}.
Besides deployable structures both big and small, the programmable shape transformations achievable with origami are amenable for driving locomotion in robotics~\cite{novelino2020untethered,wu2021stretchable,sadeghi2021tmp,bhovad2019peristaltic}, storing mechanical energy~\cite{li2016recoverable,addis2023connecting} and mechanical information~\cite{treml2018origami,jules2022delicate,bhovad2021physical}.
Further, because of the intricate connection between ``form'' and ``function'', origami is a powerful tool for developing metamaterials with tunable thermal~\cite{boatti2017origami}, mechanical~\cite{liu2022triclinic,miyazawa2021heterogeneous,misseroni2022experimental,zhai2020situ,silverberg2014using,schenk2013geometry,liu2018topological,grasinger2022multistability,pratapa2019geometric,brunck2016elastic} or electromagnetic~\cite{sessions2019origami} responses.

Techniques and algorithms have been developed for designing an origami crease pattern with particular kinds of mechanisms~\cite{gillman2018design,gillman2019discovering,shende2021bayesian,lee2024designing}, or which, upon folding, approximates a given surface~\cite{feng2020designs,feng2020helical,dudte2021additive,lang2018rigidly,dang2022inverse,dorn2022kirigami,li2020theory,dieleman2020jigsaw,liu2024design,dieleman2018origami,addis2023connecting}.
Similarly, path finding methods have been developed for planning the deployment, locomotion, etc., of an origami structure given its fold pattern, and initial and final states of interest~\cite{zhou2023low,tachi2009simulation,li2020motion}.
Underlying these tools are mathematical and computational models for the \emph{forward} problem; that is, models for the kinematics of an origami structure that answer: ``given a crease pattern, how does it fold?''.

Various computational techniques have been developed for the forward problem, each with different trade-offs in efficiency, accuracy, and complexity.
Approaches utilizing continuum solid, shell and (higher-order) frame finite elements have been proposed~\cite{fuchi2016numerical,ma2014energy,zhang2017origami}.
These methods offer a high degree of accuracy, but at greater computational expense.
Meanwhile bar-hinge-type models have emerged as alternatives that retain an engineering accuracy, but are more efficient and scalable~\cite{gillman2018truss,schenk2011origami,liu2017nonlinear,filipov2017bar,lee2022robust,chen2023multi}.
The fundamental idea of bar-hinge models is to characterize the creases as structural bar elements with both axial and torsional stiffnesses, which model the elasticity of the faces and folding about the creases, respectively.
The bar elements are then pin connected at the vertices of the origami, resulting in a Maxwell frame.
One way to reduce the dimensionality of the model is to idealize the faces of the origami structure as rigid.
This can be a good approximation when the stiffness of the faces is much larger than the torsional stiffness of the creases.
Bar-hinge type models can model rigid-facet origami by constraining the bars to be rigid~\cite{chen2018branches,mcinerney2020hidden}; and such an approach has been used for investigating the branches of folding motion off the flat state~\cite{chen2018branches} and the influence of hidden symmetries in periodic origami structures~\cite{mcinerney2020hidden}--as well as other origami problems of interest.

Assuming the faces of an origami structure are rigid leads to a condition on the fold angles of an origami structure such that it can be folded compatibly (i.e. without tearing).
Given a counterclockwise, closed path on the origami structure, which does not intersect any vertices, the \emph{loop-closure constraint} says that the product of the rotation matrices corresponding with each of the creases traversed by the path must be the identity matrix~\cite{hull2002modelling}.
Tachi~\cite{tachi2009simulation} developed a method for computational origami folding by linearizing the constraint about a given, folded state.
Then all allowable perturbations to the fold angles can be formulated as a linear system of equations.
Wu and You~\cite{wu2010modelling} proposed a model, also based on the loop-closure constraint, that utilized quaternions and dual-quaternions.
An efficient, alternative method was introduced in Zhou et al.~\cite{zhou2023low} with a redundant kinematic description, including both vertices and faces, that instead used a global constraint for ensuring vertices coincided with the corners of their respective facets.
Other notable work for the forward problem includes continuum theories which coarse grain over the fold pattern to model origami as elastic continua (e.g. plates)~\cite{zheng2022continuum,zheng2023modelling,xu2023derivation}.

While these computational methods offer many advantages, there are still several reasons that closed-form, analytical solutions are of interest.
Inverse design and optimization strategies generally rely on efficient forward models in order to be feasible.
Unsurprisingly, many of the design algorithms that are among the most efficient, and are rigorously proven to work for general problems, are based on the well-known, exact solution for degree $4$ origami vertices (i.e. vertices where $4$ creases meet), and are restricted to quadrilateral origami. 
Insight gained from analytical solutions can also be used to develop and discover new design heuristics and algorithms.
What appears to be the first analytical solution of origami kinematics was derived in a seminal paper by Huffman~\cite{huffman1976curvature}.
For a closed-path on a surface, the Gauss map produces a corresponding path on the unit sphere by taking the trace of the surface normal as it changes along the (surface) path.
Huffman appealed to the Gauss map for a degree $4$ vertex, then had the remarkable insight to use invariants of the path along with spherical trigonometry to derive implicit relationships between the fold angles.
Similar derivations, based on spherical trigonometry, were later performed by Hanna et al.~\cite{hanna2014waterbomb} to obtain an explicit solution for the folding of the symmetric $8$ fold waterbomb.
The resulting equations were later simplified~\cite{grasinger2022multistability} and generalized to other variants of the waterbomb~\cite{hanna2015force}.

An alternative approach for finding analytical solutions has roots in~\cite{hull2002modelling} where the loop-closure constraint was derived by modeling the folded state of origami using affine transformations.
There it was recognized that the deformation of the origami could be described by products of rotations.
Later work recognized the connection between these affine transformations and fundamental concepts in continuum solid mechanics such as the deformation map, the deformation gradient, and Lagrangian approaches to kinematics~\cite{feng2020designs,feng2020helical,liu2021origami,liu2024design}.
The authors used this perspective to obtain a simplified solution for degree $4$ vertices~\cite{feng2020designs,feng2020helical}; and, more importantly, \begin{inparaenum}[1)] \item demonstrated that it could be combined with concepts from objective structures~\cite{james2006objective} and group theory to efficiently generate folded states of helical origami structures~\cite{feng2020helical}, and \item developed a rigorous origami design algorithm for approximating surfaces~\cite{feng2020designs}. \end{inparaenum}
In~\cite{foschi2022explicit}, the authors use a similar, Lagrangian-like approach to derive explicit kinematic equations for general, rigid-facet degree $4$ vertices, both Euclidean and non-Euclidean (i.e. sector angles sum to $2\pi$ and do not sum to $2\pi$, respectively; non-Euclidean origami cannot be flat without facet deformation or tearing).
This work was later extended to consider various folding motions of the fully symmetric degree $6$ vertex (i.e. the degree $6$ vertex in which all of the sector angles are $\pi / 3$)~\cite{farnham2022rigid}.
Other notable analytical and rigorous results have been obtained for the hyperbolic paraboloid, or ``hypar'', origami regarding conditions for its foldability, parameterization of its surface during folding, and its bistability~\cite{demaine2011non,liu2019invariant}.

With very few exceptions, past work on obtaining exact solutions to the folding of rigid-facet origami has been restricted to single degree of freedom solutions.
Here, inspired by past work leveraging both group theory and the Lagrangian approach to origami, we derive reduced-order compatibility conditions for origami vertices with reflection and rotational symmetries.
The Lagrangian approach is outlined in \fref{sec:lagrangian}, followed by a first example in \fref{sec:miura-vertex}.
Then, using the reduced-order conditions, exact, multi degree of freedom solutions are obtained for degree $6$ and degree $8$ vertices with reflection (\fref{sec:vertices-with-reflection}) and reflection-rotation (\fref{sec:vertices-with-reflection-and-rotation}) symmetries.
We then introduce a procedure to construct lower symmetry kinematic solutions by breaking symmetry of higher order kinematic solutions in a systematic way that preserves compatibility.  
The work concludes in \fref{sec:conclusion}.

\section{The Lagrangian approach for origami kinematics} \label{sec:lagrangian}
Often in practice with origami, the object being folded is ``thin enough'' (e.g. paper) such that its bending stiffness is much less than its stiffness with respect to in-plane stretching.
Further, the bending stiffness of its facets is often much larger than the torsional stiffness of the creases.
As a result, creases represent the preferred means of deformation, and, to a good approximation, we can idealize the deformation of the structure as occurring strictly through rotations of its facets about its creases.
Because the facets are assumed rigid, one finds that the group of orthogonal linear transformations, $\OThree = \left\{\mathbf{L} \in GL\left(3\right) : \mathbf{L}^T = \mathbf{L}^{-1} \right\}$, is a powerful tool for mapping flat origami crease patterns 
into folded states (see, for example, \cite{hull2002modelling,feng2020helical,feng2020designs,liu2021origami}).
Before outlining the Lagrangian approach, we introduce the following notation:
\begin{itemize}
  \itemsep0em
  \item $\rBody$ and $\cBody$ denote the body of the origami in the reference and deformed (i.e. folded) configurations, respectively.
  \item $\xP \in \rBody$ and $\xC \in \cBody$ denotes positions in the reference and deformed configurations, respectively.
  \item $\defMap : \xP \mapsto \xC$ denotes the deformation map (or fold map), i.e. $\xC = \defMap\left(\xP\right)$.
  \item $\F$ denotes the deformation gradient, $\F = \Grad \defMap$ or $\left[\F\right]_{ij} = \partial \xCMag_i / \partial \xPMag_j$.
  \item $\refl{\unitVec} \in \OThree$ denotes a reflection about the plane which passes through the origin and is orthogonal to $\unitVec$;
  \begin{equation} \label{eq:general-reflection}
  	\refl{\unitVec} = \Iden - 2 \outProd{\unitVec}{\unitVec},
  \end{equation}
  where $\outProd{\mathbf{u}}{\mathbf{v}}$ denotes the outer product of $\mathbf{u}$ and $\mathbf{v}$ such that $\left(\outProd{\mathbf{u}}{\mathbf{v}}\right)_{ij} = u_i v_j$.
  \item $\rotAbout{\unitVec}\left(\genAngle\right) \in \SOThree \leq \OThree$ denotes a rotation of $\genAngle$ radians about the axis $\unitVec$; i.e.
  \begin{equation} \label{eq:general-rotation}
  	\rotAbout{\unitVec}\left(\genAngle\right) = \exp\left(\genAngle \skwAbout{\unitVec}\right) = \left(1 - \cos \genAngle\right) \outProd{\unitVec}{\unitVec} + \cos \genAngle \Iden + \sin \genAngle \skwAbout{\unitVec},
  \end{equation}
  where
  \begin{equation}
  	\skwAbout{\unitVec} \coloneqq \begin{pmatrix}
      0 & -\auxuj{3} & \auxuj{2} \\
      \auxuj{3} & 0 & -\auxuj{1} \\
      -\auxuj{2} & \auxuj{1} & 0
    \end{pmatrix}
  \end{equation}
  is the skew-symmetric transformation such that $\skwAbout{\mathbf{u}} \mathbf{v} = \mathbf{u} \times \mathbf{v}$ for all $\mathbf{v} \in \Reals^3$, and $\SOThree = \left\{\mathbf{L} \in \OThree : \det \mathbf{L} = 1 \right\}$ is the group of special orthogonal transformations.
\end{itemize}

In this work, we will restrict our attention to a single origami vertex.
Towards describing a folded deformation, start with any crease, label it $1$, and then proceed around the vertex in the counterclockwise direction, labeling each subsequent crease, $2, 3, \dots, \NCreases$, where $\NCreases$ is the number of creases at the vertex.
Then let the unit vector which begins at the vertex and extends along crease $i$ in the reference  configuration be denoted by $\CreaseVec_i$.
The reference configuration is chosen to be the flat state throughout.
Further, $\foldAngle_i \in \left[-\pi, \pi\right]$ denotes the supplement of the dihedral angle between the faces adjacent to crease $i$ and is referred to as the \emph{fold angle} at crease $i$.
For the facets, let the facet which is bounded by the creases $i$ and $i+1$ be denoted by $\Facet{i}$ and $\facet{i}$ in the reference configuration and deformed configuration, respectively, for $i = 1, \dots, \NCreases-1$; similarly, $\Facet{\NCreases}$ and $\facet{\NCreases}$ is the facet bounded by the creases $\NCreases$ and $1$.
As is often the case in solid mechanics, we are only interested in the deformation map up to rigid body rotations and translations so without loss of generality assume that $\facet{\NCreases} = \Facet{\NCreases}$ is fixed.
Then it is easy to see that $\xC = \left(\rotAbout{\CreaseVec_1}\left(\foldAngle_1\right)\right) \xP$ for $\xP \in \Facet{1}$.
Similarly, because by assumption there are no tears and thus $\facet{1}$ is joined to $\facet{2}$ at crease $2$, we have that $\xC = \left(\rotAbout{\CreaseVec_1}\left(\foldAngle_1\right)\right) \left(\rotAbout{\CreaseVec_2}\left(\foldAngle_2\right)\right) \xP$ for $\xP \in \Facet{2}$.
Generalizing this argument to all facets leads to the following result,
\begin{equation} \label{eq:fold-map}
	\xC = \defMap\left(\xP\right) = \left(\prod_{i=1}^{j} \rotAbout{\CreaseVec_i}\left(\foldAngle_i\right)\right) \xP, \quad \xP \in \Facet{j},
\end{equation}
As $\defMap\left(\xP\right)$ is  a linear transformation of $\xP$, the deformation gradient is simply
\begin{equation}
	\F\left(\xP\right) = \prod_{i=1}^j \rotAbout{\CreaseVec_i}\left(\foldAngle_i\right), \quad \xP \in \Facet{j},
\end{equation}
which is piecewise constant.
For brevity, let $\F_i$ be defined such that $\facet{i} = \F_i \Facet{i}$.

Recall that the above formulation of the deformation map assumes that \begin{inparaenum}[1)] \item $\Facet{\NCreases}$ is fixed and \item there are no tears, i.e. the deformation is compatible. \end{inparaenum}
As a result, a necessary and sufficient condition~\footnote{
	The condition is only necessary and no longer sufficient when facets are prohibited from passing through each other~\cite{hull2002modelling}; i.e., satisfying \eqref{eq:compatibility} alone does not guarantee that the origami vertex will not contact or intersect itself.
} for a compatible deformation is given by the loop-closure constraint~\cite{hull2002modelling}
\begin{equation} \label{eq:compatibility}
  \F_{\NCreases}\left(\foldAngle_1, \dots, \foldAngle_{\NCreases}\right) = \prod_{i=1}^{\NCreases} \rotAbout{\CreaseVec_i}\left(\foldAngle_i\right) = \Iden.
\end{equation}
Choosing the deformation map from among the group of rigid body translations and rotations\footnote{Here, by ``rigid rotation'', we mean that the entire origami is rotated such that all of the fold angles are preserved.} of the origami such that $\F_{\NCreases} = \Iden$ is a standard convention which is convenient for deriving \eqref{eq:compatibility}.
However, in the proceeding sections we will adjust this convention based on the symmetry of the underlying crease pattern in order to simplify the reduced order compatibility conditions.
For example, when considering vertices with reflection symmetry about the plane which cuts through crease $1$, it will be convenient to adopt the convention
\begin{equation}
\begin{split}
\F_1 = \rotate{\CreaseVec_1}{\frac{\foldAngle_1}{2}}, \quad \F_2 = \rotate{\CreaseVec_1}{\frac{\foldAngle_1}{2}} \rotate{\CreaseVec_2}{\foldAngle_2}, \quad \dots, \\
\quad \F_{\NCreases} = \rotate{\CreaseVec_1}{\frac{\foldAngle_1}{2}} \prod_{i=2}^{\NCreases} \rotate{\CreaseVec_{i}}{\foldAngle_{i}} = \rotate{\CreaseVec_1}{-\frac{\foldAngle_1}{2}}
\end{split}
\end{equation}
which corresponds to rigidly rotating the body by $\rotate{\CreaseVec_1}{-\foldAngle_1 / 2}$ and where the last equality corresponds to multiplying both sides of \eqref{eq:compatibility} by $\rotate{\CreaseVec_1}{-\foldAngle_1 / 2}$.
The exact choice of the deformation map will be made clear in each subsequent example provided.

While \eqref{eq:compatibility} is quite general, it is also highly nonlinear and difficult to solve except in special cases.
In this work, we focus on vertices with various symmetries and find that the Lagrangian description facilitates leveraging those symmetries to reduce the complexity of the problem and obtain closed-form solutions.

\section{A first example: $4$-vertex with reflection symmetry} \label{sec:miura-vertex}

\begin{figure}
	\centering
	\includegraphics[width=\linewidth]{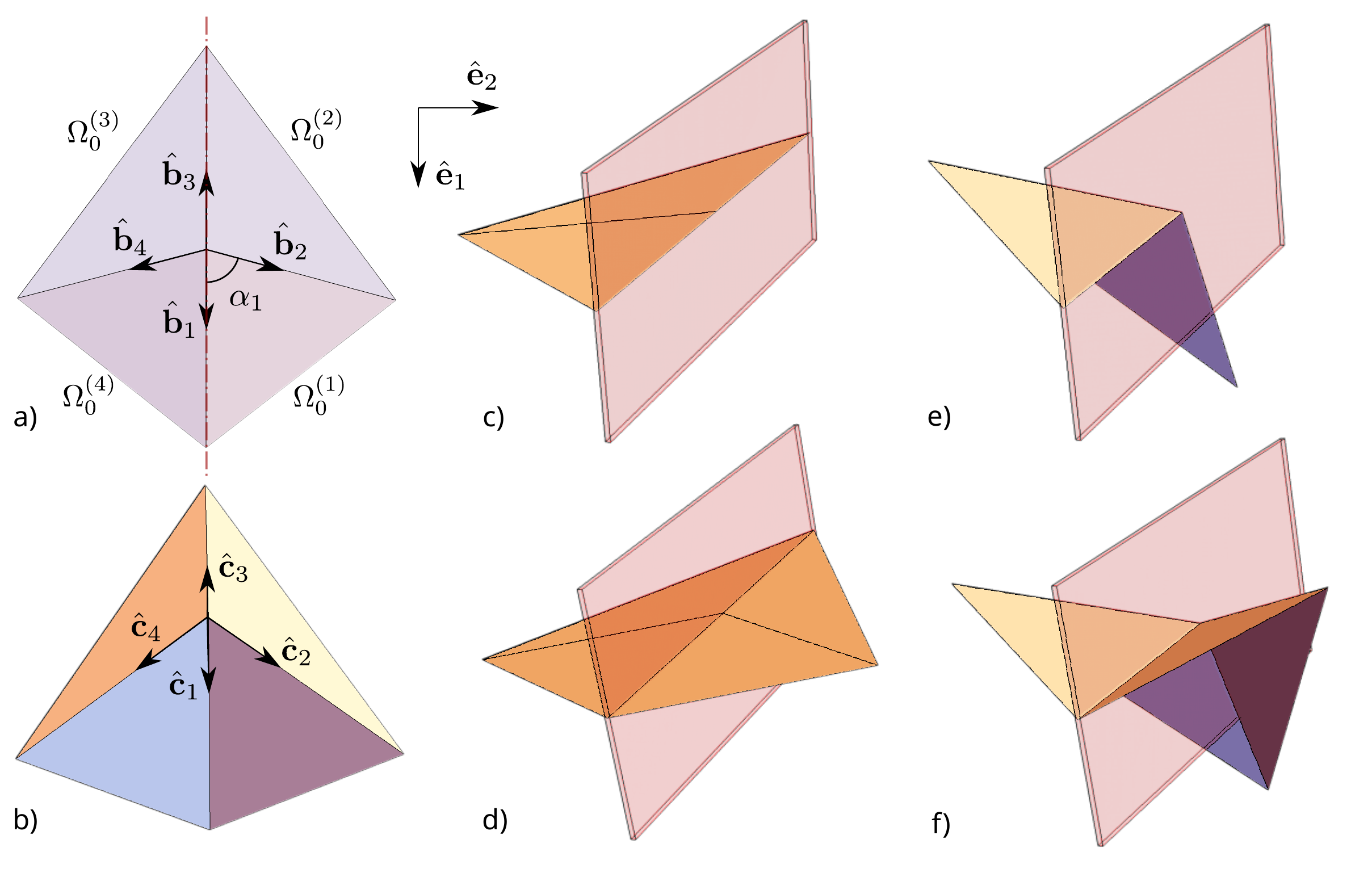}
	\caption{
		\textbf{Geometry, coordinate system, and symmetric folding of the degree $4$ vertex with reflection symmetry.}
		a) Geometry of the $4$ vertex with a reflection symmetry and parameterized by the sector angle between the first two creases, $\sectAngle_1$.
		b) Folded configuration with crease vectors $\creaseVec_i, i = 1, 2, 3, 4$ superimposed.
		c) Half of the flat configuration.
		d) Full flat configuration generated by the action $\refl{\etwo} \left(\Facet{1} \cup \Facet{2}\right)$.
		e) Half of the flat configuration folded such that $\creaseVec_3$ is in the plane of symmetry, i.e. $\creaseVec_3 \cdot \etwo = 0$.
		f) Full folded configuration generated by the action $\refl{\etwo} \left(\facet{1} \cup \facet{2}\right)$.
	}
	\label{fig:4-vertex}
\end{figure}

We introduce the key ideas of the overarching approach by using degree $4$ vertices as a first example.
The $4$ vertex has a single degree of freedom, but two separate kinematic folding solutions that bifurcate off of the flat configuration.
The kinematics of degree $4$ vertices are well-known~\cite{huffman1976curvature,foschi2022explicit}, so it presents a natural starting point for illustrating the approach proposed herein.
The $4$ vertices considered here have two collinear folds and a reflection symmetry about the axis which contains the collinear folds.
This vertex is parameterized by a single parameter: the sector angle between the first two creases, $\sectAngle_1$.
Let the coordinate system be such that the origin is at the vertex, the reflection symmetry is orthogonal to the $\etwo$ direction, and the $\ethree$ direction is orthogonal to the flat configuration.
The geometry of the vertex and choice of coordinate system are shown in \fref{fig:4-vertex}.

\newcommand{\auxset}{\generic}
An important observation from \fref{fig:4-vertex} is that the flat state can be constructed by starting with $\Facet{1}$ and $\Facet{2}$ (the facets between creases $1$ and $2$, and $2$ and $3$, respectively), and then joining the facets with a copy of themselves reflected about the plane orthogonal to $\etwo$.
Let $\Facet{h} \coloneqq \Facet{1} \bigcup \Facet{2}$.
Then $\rBody = \orbit{\Facet{h}}{\group}$ where $\group = \set{\Iden, \refl{\etwo}}$ and $\orbit{\auxset}{\otherGroup} \coloneqq \bigcup_{g \in \otherGroup} \; g \left(\auxset\right)$ is called the orbit of $\auxset$ under the group $\otherGroup$.
The implication here is that we can generate a folded configuration by folding $\Facet{h}$ (i.e. $\Facet{h} \rightarrow \facet{h}$) and then letting $\cBody = \orbit{\facet{h}}{\group}$.
The caveat is that we require that $\creaseVec_{1} \cdot \etwo = \creaseVec_{3} \cdot \etwo = 0$ for there to be no tears, where $\creaseVec_{i}$ is the unit vector along crease $i$ in the deformed configuration.
In other words, the orbit will be a compatible configuration provided that $\facet{h}$ and $\refl{\etwo} \facet{h}$ overlap at creases $1$ and $3$.
This represents a dimensional reduction of \eqref{eq:compatibility}, which depends on $\foldAngle_1, \dots, \foldAngle_4$, to the condition that $\creaseVec_{3} \cdot \etwo = 0$, which only depends on $\foldAngle_1$ and $\foldAngle_2$.
As alluded to previously, we recall that the deformation map is only meaningful up to rigid rotations and translations of the entire body and so we first rotate $\Facet{1}$ and $\Facet{2}$ by $\rotAbout{\CreaseVec_1}\left(\foldAngle_1 / 2\right)$.
As a result, our deformation map has been adjusted such that
\begin{equation}
	\F_1 = \rotAbout{\CreaseVec_1}\left(\frac{\foldAngle_1}{2}\right), \quad \F_2 = \rotAbout{\CreaseVec_1}\left(\frac{\foldAngle_1}{2}\right) \rotAbout{\CreaseVec_2}\left(\foldAngle_2\right).
\end{equation}
Given this adjusted but equivalent formulation, we now solve for $\foldAngle_2$ as a function of $\foldAngle_1$ such that
\begin{equation} \label{eq:reduced-4-vertex}
	\creaseVec_3 \cdot \etwo = \left(\F_2 \CreaseVec_3\right) \cdot \etwo = \left(\rotAbout{\CreaseVec_1}\left(\frac{\foldAngle_1}{2}\right) \rotAbout{\CreaseVec_2}\left(\foldAngle_2\right) \CreaseVec_3 \right) \cdot \etwo = 0.
\end{equation}
A schematic of this solution process is shown in \fref{fig:4-vertex}.
The solution to \eqref{eq:reduced-4-vertex} is readily obtained as
\begin{equation} \label{eq:symmetric-4-vertex} 
	\foldAngle_1 = 2 \arctan \left(\cos \sectAngle_1 \csc \foldAngle_2 \left(\cos\foldAngle_2 -1\right)\right)
\end{equation}
which agrees with solutions for $4$ vertices published in the literature~\cite{huffman1976curvature,feng2020helical,dieleman2018origami,foschi2022explicit}.
From here we have that $\foldAngle_4 = \foldAngle_2$ (since $\CreaseVec_4 = \refl{\etwo} \CreaseVec_2$) and it remains to compute $\foldAngle_3$.
In principle, we could substitute $\foldAngle_1, \foldAngle_2,$ and $\foldAngle_4$ into \eqref{eq:compatibility} and solve for $\foldAngle_3$.
However, once again, the Lagrangian perspective lends itself towards the solution.
We instead use what is currently known about the deformation map:
\begin{equation} \label{eq:phi3}
\begin{split}
	\unNormalVec_{2} &= \creaseVec_2 \times \creaseVec_3, \\
	\foldAngle_3 &= -\sgn\left(\unNormalVec_2 \cdot \etwo\right) \arccos \left(\frac{\unNormalVec_2 \cdot \left(\refl{\etwo} \unNormalVec_2\right)}{\unNormalVec_2 \cdot \unNormalVec_2}\right).
\end{split}
\end{equation}
where $\unNormalVec_2$ is a vector orthogonal to $\facet{2}$\footnote{A simplified solution to \eqref{eq:phi3} is given in \fref{app:algebraic-4-vertex}.}.

\section{Vertices with a reflection symmetry} \label{sec:vertices-with-reflection}

While the previous section appealed to a geometric lens, here we explore the method through an algebraic lens, and its generalization to even number of creases, $\NCreases$.
In general, for an origami which can be described by $\rBody = \orbit{\Facet{c}}{\group}$ for some unit cell $\Facet{c}$ and group $\group$, we can rewrite \eqref{eq:compatibility} as
\begin{equation} \label{eq:group-compatibility}
	\prod_{i=1}^{\NCreases} \rotAbout{g_i \cdot \CreaseVec_i'} \left(\foldAngle_i\right) = \Iden
\end{equation}
where $g_i \cdot \CreaseVec_i' = \CreaseVec_i$ for some $g_i \in \group$ and $\CreaseVec_i' \in \Facet{c}$ and where the index $i$ is necessary because, in general, not all creases will be mapped from the same crease in $\Facet{c}$ or under the same action transformation, $g_i$.
The algebraic difficulty of \eqref{eq:group-compatibility} stems from the fact that \begin{inparaenum}[1)] \item the rotations are taken about different axes and \item the transformations do not commute. \end{inparaenum}
Thus, a potential strategy for simplifying \eqref{eq:group-compatibility} consists of reformulating in terms of like axes and taking advantage of transformations that commute in order to combine terms.
Towards this end, we note the following useful identities
\begin{subequations}
	\label{eq:identities}
	\begin{align}
		\label{eq:rotations-inverse}
		\rotAbout{\unitVec}^T\left(\genAngle\right) &= \rotate{\unitVec}{-\genAngle} = \rotate{-\unitVec}{\genAngle}, \\
		\label{eq:reflections-inverse}
		\refl{\unitVec} &= \refl{\unitVec}^T \\
		\label{eq:rotations-about-same-axis}
		\rotate{\unitVec}{\genAngle} \rotate{\unitVec}{\genAngle'} &= \rotate{\unitVec}{\genAngle + \genAngle'}, \\
		\label{eq:rotations-about-transformed-axis}
		\rotate{\orthoTrans \unitVec}{\genAngle} &= \orthoTrans \rotate{\unitVec}{\left|\orthoTrans\right| \genAngle} \orthoTrans^T, \quad \qquad \forall \orthoTrans \in \OThree, \\
		\label{eq:rotation-reflection-about-ortho-axes}
		\rotAbout{\unitVec}\left(\genAngle\right) = \rotAbout{\refl{\otherUnitVec}\unitVec}\left(\genAngle\right) &= \refl{\otherUnitVec} \rotAbout{\unitVec}\left(-\genAngle\right) \refl{\otherUnitVec}, \qquad \forall \unitVec, \otherUnitVec \text{ such that } \unitVec \cdot \otherUnitVec = 0.\\
		\label{eq:two-reflections-is-a-rotation}
		\refl{\unitVec} \refl{\otherUnitVec} &= \rotate{\hat{\mathbf{w}}}{2 \arccos \left(\unitVec \cdot \otherUnitVec\right)}, \quad \quad \hat{\mathbf{w}} \coloneqq \frac{\unitVec \times \otherUnitVec}{\left|\unitVec \times \otherUnitVec\right|}
		\\ 
		\label{eq:perpendicular-reflections-commute}
		\refl{\unitVec} \refl{\otherUnitVec} &= \refl{\otherUnitVec} \refl{\unitVec}, \qquad \qquad \qquad \forall \unitVec, \otherUnitVec \text{ such that } \unitVec \cdot \otherUnitVec = 0, \\
		\label{eq:rotation-reflection-about-same-axis}
		\refl{\unitVec} \rotate{\unitVec}{\genAngle} &= \rotate{\unitVec}{\genAngle} \refl{\unitVec}.
	\end{align}
\end{subequations}

Consider a vertex with crease unit vectors \begin{equation} \CreaseVec_1, \; \dots, \;  \CreaseVec_{\NCreases/2}, \quad \CreaseVec_{\NCreases / 2 + 1} = -\CreaseVec_1, \quad \CreaseVec_{\NCreases / 2 + 2} = \refl{\etwo} \CreaseVec_{\NCreases/2}, \quad \dots, \quad \CreaseVec_{\NCreases} = \refl{\etwo} \CreaseVec_{2}.
\end{equation}
such that the vertex has a reflection symmetry in the flat, reference configuration.
The coordinate system is chosen as Euclidean such that $\ethree$ is orthogonal to the flat configuration and the reflection symmetry is about the plane orthogonal to $\etwo$.
Let $\group = \set{\Iden, \refl{\etwo}}$.
Then \eqref{eq:group-compatibility} takes the form of
\begin{equation}
	\rotate{\CreaseVec_1}{\foldAngle_1} \: \dots \: \rotate{\CreaseVec_{\NCreases/2}}{\foldAngle_{\NCreases/2}} \rotate{-\CreaseVec_1}{\foldAngle_{\NCreases/2 + 1}} \rotate{\refl{\etwo} \CreaseVec_{\NCreases/2}}{\foldAngle_{\NCreases/2+2}} \: \dots \: \rotate{\refl{\etwo} \CreaseVec_{2}}{\foldAngle_{\NCreases}} = \Iden,
\end{equation}
which by \eqref{eq:identities} is equivalent to
\begin{equation} \label{eq:refl-compat}
\begin{split}
&\rotate{\CreaseVec_1}{\frac{\foldAngle_1}{2}} \: \dots \: \rotate{\CreaseVec_{\NCreases/2}}{\foldAngle_{\NCreases/2}} \rotate{\CreaseVec_1}{-\foldAngle_{\NCreases/2+1}} \times \\ &\refl{\etwo} \rotate{\CreaseVec_{\NCreases/2}}{-\foldAngle_{\NCreases/2+1}} \dots \rotate{\CreaseVec_{2}}{-\foldAngle_{\NCreases}} \rotate{\CreaseVec_1}{-\frac{\foldAngle_1}{2}} \refl{\etwo} = \Iden.
\end{split}
\end{equation}
Similar to \eqref{eq:consistency-4-vertex-23}, we find that
\begin{equation} \label{eq:general-reflection-consistency-condition-full}
	\rotate{\CreaseVec_1}{\frac{\foldAngle_1}{2}} \dots \rotate{\CreaseVec_{\NCreases/2}}{\foldAngle_{\NCreases/2}} \rotate{\CreaseVec_1}{-\foldAngle_{\NCreases/2+1}} = \refl{\etwo} \rotate{\CreaseVec_1}{\frac{\foldAngle_1}{2}} \dots \rotate{\CreaseVec_{\NCreases/2}}{\foldAngle_{\NCreases/2}} \refl{\etwo},
\end{equation}
represents a reduced-order formulation for solutions of \eqref{eq:group-compatibility} which have a symmetry described by $\group$.
This can be reformulated as
\begin{equation} \label{eq:general-reflection-consistency-condition}
	\left(\rotate{\CreaseVec_1}{\foldAngle_1 / 2} \rotate{\CreaseVec_2}{\foldAngle_2} \dots \rotate{\CreaseVec_{\NCreases/2}}{\foldAngle_{\NCreases/2}} \CreaseVec_{\NCreases/2+1}\right) \cdot \etwo = 0.
\end{equation}
Upon satisfying \eqref{eq:general-reflection-consistency-condition}, and plugging into \eqref{eq:refl-compat}, one readily sees that
\begin{equation}
	\phi_{\NCreases / 2 + 2} = \phi_{\NCreases / 2}, \quad \phi_{\NCreases / 2 + 3} = \phi_{\NCreases / 2 - 1}, \quad \dots, \quad \phi_{\NCreases} = \phi_{2}
\end{equation}
is a solution.
An example of the algebraic approach applied to degree $4$ vertices with reflection symmetry is given in \fref{app:algebraic-4-vertex}.
While the primary focus of this work is on closed-form solutions, a reduced order computational approach, based on \eqref{eq:general-reflection-consistency-condition} and seminal work by Tachi~\cite{tachi2009simulation}, is also outlined in the supplementary information. 

We will next obtain multi-degree of freedom, closed-form solutions to \eqref{eq:general-reflection-consistency-condition} for degree $6$ and degree $8$ vertices.
In general, equation \eqref{eq:general-reflection-consistency-condition} is still difficult to solve because it is nonlinear and contains the unknowns as arguments of trigonometric functions.
However, in comparison to \eqref{eq:compatibility}, there multiple simplifications worth noting.
The first, which has already been mentioned, is that the number of unknowns has been reduced--specifically, the number of degrees of freedom is reduced from $\NCreases - 3$ to $\NCreases / 2 - 1$.
The second simplification is also important: \emph{while \eqref{eq:compatibility} represents a system of $3$ simultaneous, nonlinear equations, \eqref{eq:general-reflection-rotation-consistency-condition} is only a single nonlinear equation instead of a system.} 
Throughout this work, kinematic solutions are found and simplified using the powerful symbolic computation algorithms in Mathematica~\cite{Mathematica}.
Although simultaneous nonlinear equations can be prohibitively difficult to solve, Mathematica has powerful algorithms for using a wide array of trigonometric identities to solve and simplify realizations of \eqref{eq:general-reflection-rotation-consistency-condition}.

\subsection{$6$-vertex}

\begin{figure}
	\centering
	\includegraphics[width=0.60\linewidth]{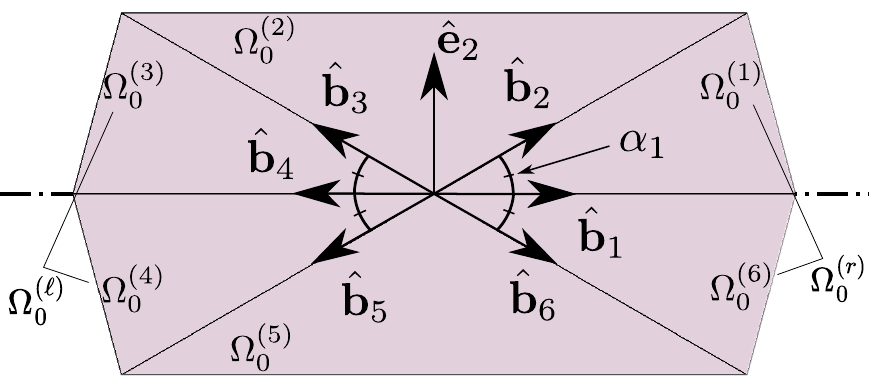}
	\caption{
		\textbf{Geometry, coordinate system, and symmetry of the degree $6$ vertex with reflection symmetry.}
		a) Geometry of the $6$ vertex with a reflection symmetry.
		The vertex is parameterized by the sector angle between the first two creases, $\sectAngle_1$.
		b) Coordinate system, axis of symmetry, and facet labels, $\Facet{i}, i = 1, \dots, 6$. 
            We define $\Facet{r} \coloneqq \Facet{1} \bigcup \Facet{6}$ and $\Facet{\ell} \coloneqq \Facet{3} \bigcup \Facet{4}$.
	}
	\label{fig:setup-6-vertex}
\end{figure}

In this section, we consider $6$ vertices with a reflection symmetry parameterized by $\sectAngle_1$ such that $\sectAngle_3 = \sectAngle_1$ and $\sectAngle_2 = \pi - 2 \sectAngle_1$, as shown in \fref{fig:setup-6-vertex}.
The reduced order compatibility condition \eqref{eq:general-reflection-consistency-condition-full} for this system admits a closed-form solution:
\begin{hlbreakable}
\begin{equation} \label{eq:phi1-6-fold}
	\begin{split}
		\foldAngle_1 &= \convToFoldAngle\left(2 \arctan\left(-\argSign_6 \numer_6, \argSign_6 \denom_6\right)\right), \\
		\numer_6 &= \cCos \sectAngle_1 \cSin \sectAngle_1 \Bigg(
		2 \cCos^2 \sectAngle_1 \cSin^2 \frac{\foldAngle_3}{2} \left(1 - 2 \cCos \foldAngle_2\right) - \cSin^2 \sectAngle_1 + \\
		& \qquad \qquad \qquad \qquad \quad \cCos \foldAngle_3 \left(1 - \cCos\foldAngle_2 + \cSin^2 \sectAngle_1\right) +
		\cSin \foldAngle_2 \cSin \foldAngle_3
		\Bigg), \\
		\denom_6 &= \cSin \sectAngle_1 \left(
		\cSin \foldAngle_2 +
		2 \cCos \left(2 \sectAngle_1\right) \cSin \foldAngle_2 \cSin^2 \frac{\foldAngle_3}{2} +
		\cCos \foldAngle_2 \cSin \foldAngle_3 \right), \\
            \argSign_6 &= \pm 1,
	\end{split}
\end{equation}
where $\arctan\left(y, x\right)$ considers the quadrant in which  $\left(x, y\right)$ resides, and where $\argSign_6$ helps specify the quadrant of $\left(\denom_6, \numer_6\right)$.
Of these two possible solutions, only one satisfies compatibility for a general $\foldAngle_2, \foldAngle_3,$ and $\sectAngle_1$.
While it is difficult to determine analytically which solution is valid, we speculate that
\begin{equation}
		\argSign_6 = \sgn \left(
		\cCos \foldAngle_2 \cSin \foldAngle_3 +
		2 \cCos \left(2 \sectAngle_1\right) \cSin \foldAngle_2 \cSin^2 \frac{\foldAngle_3}{2} +
		\cSin \foldAngle_2 \right)
\end{equation}
is correct, and suggest relying on numerical verification in practice.
\end{hlbreakable}
Lastly,  
\begin{equation}
	\convToFoldAngle\left(\genAngle\right) = \begin{cases}
		\convToFoldAngle\left(\genAngle + 2\pi\right) & \genAngle < -\pi \\
		\convToFoldAngle\left(\genAngle - 2\pi\right) & \genAngle > \pi \\
		\genAngle & \text{otherwise}
	\end{cases},
\end{equation}
converts fold angles into the allowable range prior to contact of adjacent faces (i.e. $\left[-\pi, \pi\right]$).
The solution was found and simplified using the powerful symbolic computation algorithms in Mathematica~\cite{Mathematica}.
The Mathematica notebook can be found \href{https://github.com/grasingerm/symmetric-Lagrangian-origami} {here}.
It was then verified by numerically evaluating $\lVert \prod_{i=1}^6 \rotAbout{\CreaseVec_i}\left(\foldAngle_i\right) - \Iden \rVert_2$ and finding that it was $\lessapprox 10^{-16}$ throughout the domain, $\left(\foldAngle_{2}, \foldAngle_{3}\right) \in \left[-\pi, \pi\right] \times \left[-\pi, \pi\right]$.

\begin{figure}
	\centering
	\includegraphics[width=\linewidth]{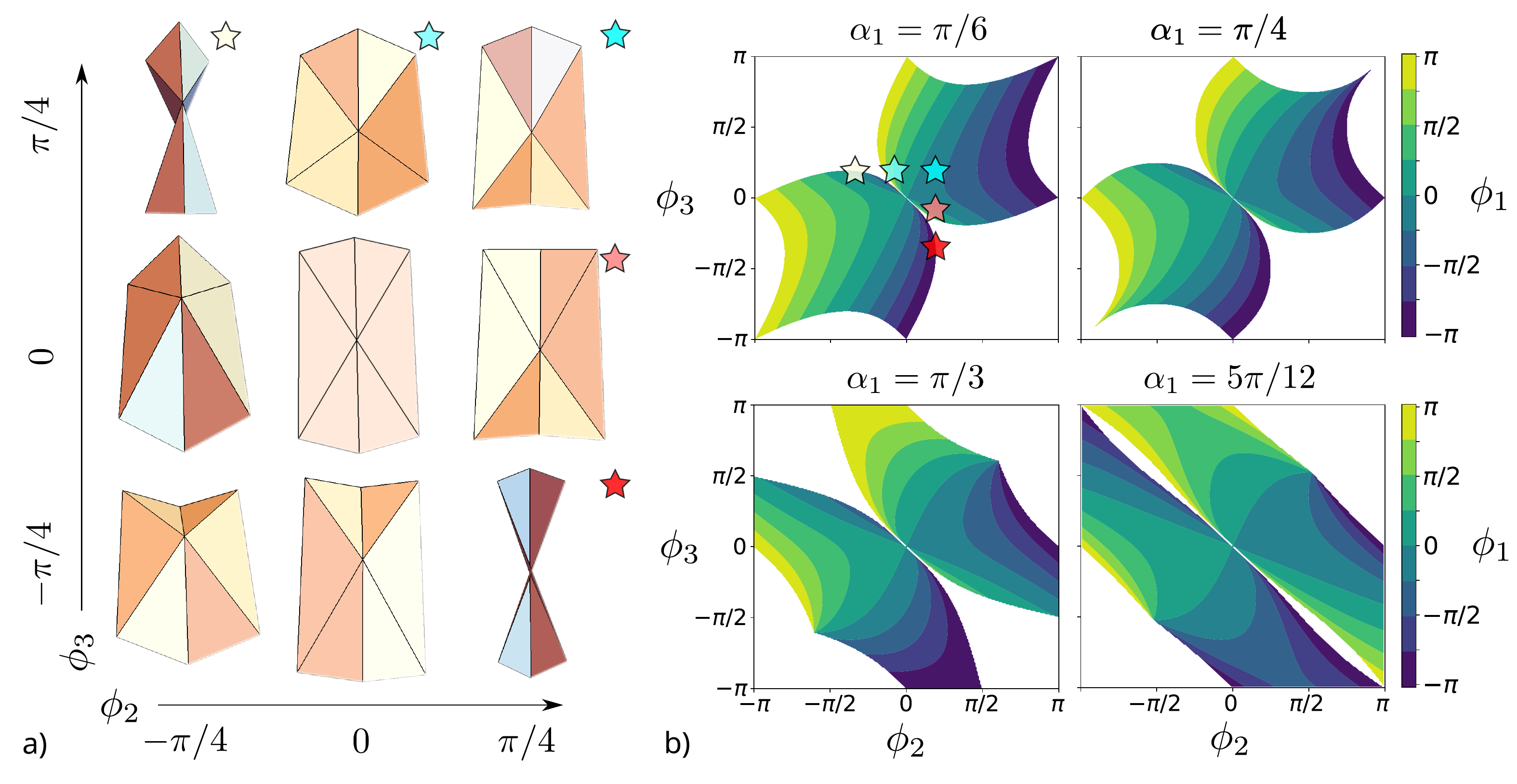}
	\caption{
		\textbf{Degrees of freedom and kinematics of the degree $6$ vertex with reflection symmetry.}
		a) Folded states of a 6-vertex with a reflection symmetry and with $\sectAngle_1 = \pi / 6$ and $\sectAngle_2 = \pi - 2 \sectAngle_1$.
		Traversing columns (i.e. right-to-left) corresponds with changing $\foldAngle_2$ and traversing rows (i.e. bottom-to-top) corresponds with changing $\foldAngle_3$.
		b) Contours of $\foldAngle_1$ as a function of $\foldAngle_2$ (x-axis) and $\foldAngle_3$ (y-axis) for $\sectAngle_1 = \pi / 6, \pi / 4, \pi / 3,$ and $5 \pi / 12$ (top-left, top-right, bottom-left, and bottom-right, respectively).
        White regions are inadmissible due to contact; note that the number of holes in admissible space, the size of the admissible regions, and the connectivity between admissible regions changes with $\sectAngle_1$.
	}
	\label{fig:6-dofs-and-contours}
\end{figure}

Folded states of the $6$-vertex with $\sectAngle_1 = \pi / 6$ for various combinations of $\foldAngle_2 \in \set{-\pi / 4, 0, \pi / 4}$ and $\foldAngle_3 \in \set{-\pi / 4, 0, \pi / 4}$ are shown in \fref{fig:6-dofs-and-contours}.a; and contours of $\foldAngle_1$ as a function of $\foldAngle_2$ and $\foldAngle_3$ are shown in \fref{fig:6-dofs-and-contours}.b.
Regions of $\left(\foldAngle_2, \foldAngle_3\right)$ space which are inadmissible due to self-contact are shown in white.
The procedure for detecting self-contact is outlined in \fref{app:self-contact}.
An animation of the highlighted folding trajectory is provided in the ESI (see Animation \#1).
When self contact occurs in any of the folding animations, the frame is tinted red to signal the configuration is inadmissible.

\begin{figure}
	\centering
	\includegraphics[width=\linewidth]{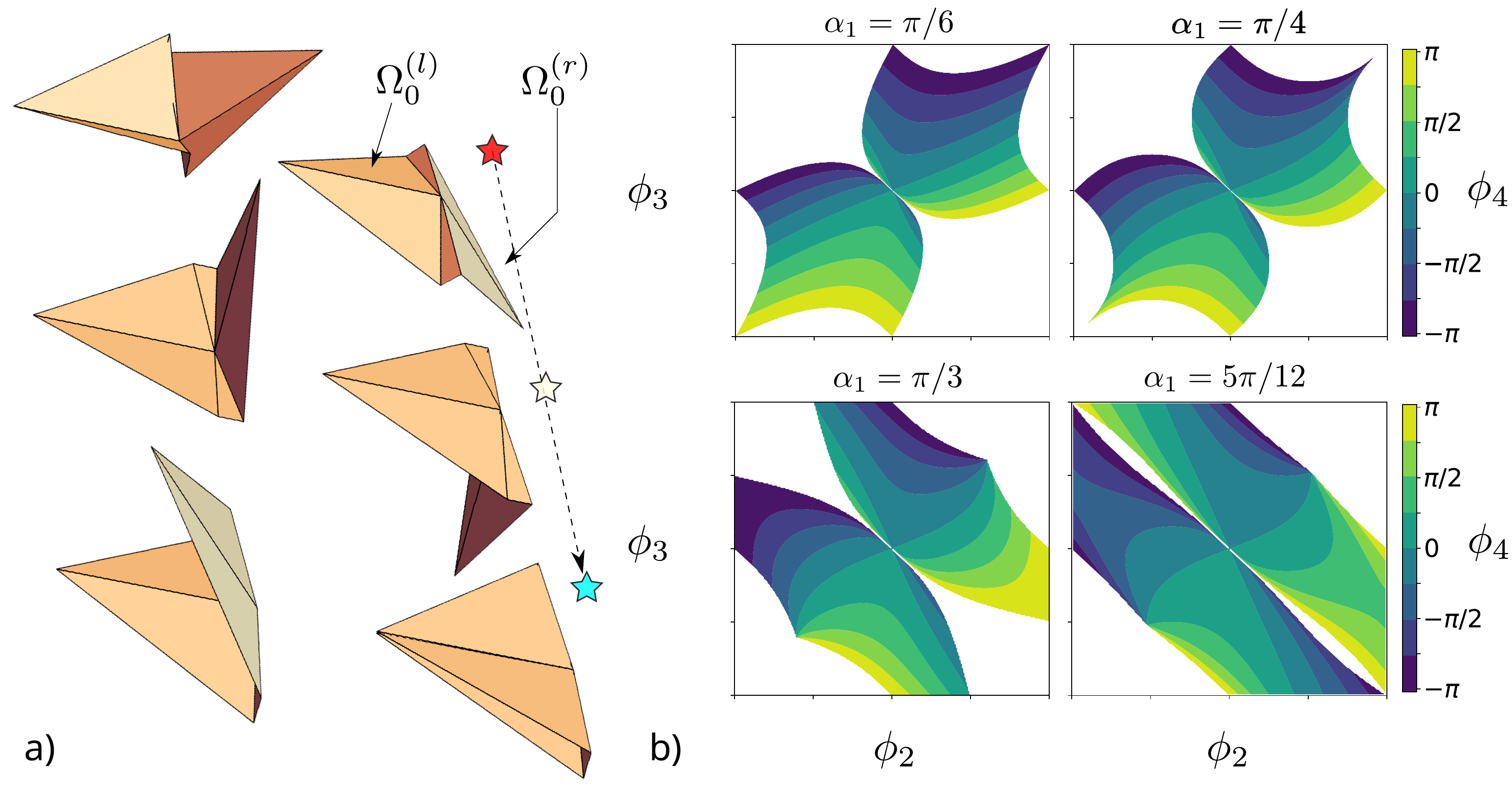}
	\caption{
		\textbf{Dependence of $6$ vertex kinematics on $\sectAngle_1$ and hinge-like behavior.}
		a) Folded states of a 6-vertex with a reflection symmetry and with $\sectAngle_1 = 11 \pi / 24$ and $\sectAngle_2 = \pi - 2 \sectAngle_1$.
		As $\sectAngle_1 \rightarrow \pi / 2$, creases $2, 3, 5$ and $6$ act collectively as a kind of hinge such that $\facet{r}$ and $\facet{\ell}$ can fold toward or away from each other without a lot of folding across their middle creases (crease $1$ and $4$, respectively).
		b) Contours of $\foldAngle_4$ as a function of $\foldAngle_2$ (x-axis) and $\foldAngle_3$ (y-axis) for $\sectAngle_1 = \pi / 6, \pi / 4, \pi / 3,$ and $5 \pi / 12$ (top-left, top-right, bottom-left, and bottom-right, respectively).
	}
	\label{fig:6-dofs-and-contours_Phi4}
\end{figure}

The top row of kinematic contour plots in \fref{fig:6-dofs-and-contours}.b corresponds with vertices where the folds can be divided into two groups of creases which are ``near'' to each other (e.g. when $\sectAngle_1 = \pi / 12$ (top-left), creases 1, 2, and 6 are all within $\pi / 6$ of each other and similarly, creases 3, 4, and 5 are all within $\pi / 6$ of each other).
The way that crease 1 folds as a function of $\foldAngle_2$ and $\foldAngle_3$ is similar in character for each of these two cases.
In contrast, the bottom row of \fref{fig:6-dofs-and-contours}.b shows that, $\foldAngle_1 = \foldAngle_1\left(\foldAngle_2, \foldAngle_3\right)$ is characteristically different when increasing $\sectAngle_1$ such that creases $2$ and $3$ are closer together.
The kinematic contours for $\sectAngle_1 = \pi / 3$ and $5 \pi / 12$ are similar, but the $\sectAngle_1 = 5 \pi / 12$ contours for moderate $\left|\foldAngle_1\right|$ have expanded relative to their $\sectAngle_1 = \pi / 3$ counterparts; vice versa, the contours for greater magnitude fold angles (i.e. $\left|\foldAngle_1\right| > \pi / 2$) have contracted relative to their analog in the $\sectAngle_1 = \pi / 3$ case.
Crease $1$ folds less, in general, as a function of $\foldAngle_2$ and $\foldAngle_3$, for increasing $\sectAngle_1$ (when $\sectAngle_1 \geq \pi / 3$).
This phenomena is illustrated in \fref{fig:6-dofs-and-contours_Phi4}.a where various folded states of a 6-vertex with $\sectAngle_1 = 11 \pi / 24$ are shown.
Let $\facet{r} \coloneqq \facet{1} \bigcup \facet{6}$ and $\facet{\ell} \coloneqq \facet{3} \bigcup \facet{4}$ (see \fref{fig:setup-6-vertex}).
As $\sectAngle_1 \rightarrow \pi / 2$, creases $2$, $3$, $5$ and $6$ act collectively as a kind of hinge such that $\facet{r}$ and $\facet{\ell}$ can fold toward or away from each other without much flexing across their middle creases. 
The change in geometry and kinematics as a function of $\sectAngle_1$ has important implications for self-contact as well. 
The number of holes in admissible space, the size of the admissible regions, and the connectivity between admissible regions changes with $\sectAngle_1$.
For instance, the bottom-left and top-right admissible regions only connect at a single point, the flat configuration (i.e. $\left(0, 0\right)$), for the $\sectAngle_1 = \pi / 6$ and $\pi / 4$ cases, but at $3$ points for the $\sectAngle = 5 \pi / 12$ case.
To help illustrate why, animations of the $\sectAngle_1 = \pi / 4$ and $\sectAngle_1 = 5 \pi / 12$ vertices folding along $\left(\foldAngle_2 = 0 \rightarrow -\pi, \foldAngle_3 = \pi\right)$ and then $\left(\foldAngle_2 = -\pi, \foldAngle_3 = \pi \rightarrow 0\right)$ are provided in the ESI as Animation \#2 and Animation \#3, respectively.
When $\sectAngle_1$ is small, facets $\Facet{2}$ and $\Facet{5}$ are ``large'' and appear to contact each other through nearly the entire folding motion of the $\sectAngle_1 = \pi / 4$ vertex.
Whereas $\Facet{2}$ and $\Facet{5}$ are much smaller when $\sectAngle_1 = 5 \pi / 12$, so that the two flaps on either end of the vertex are able to nest within each other during the folding.

\newcommand{\auxc}{c_{\sectAngle}}
By construction, $\foldAngle_5 = \foldAngle_3$ and $\foldAngle_6 = \foldAngle_2$.
Having obtained $\foldAngle_1 = \foldAngle_1\left(\foldAngle_2, \foldAngle_3\right)$ (equation \eqref{eq:phi1-6-fold}), what remains of the kinematic description of the 6-fold vertex is $\foldAngle_4$.
Similar to \fref{sec:miura-vertex}, we derive the remaining fold angle by appealing to the deformation map and noting that
\begin{equation}
\begin{split}
	\creaseVec_3 &= \F_2 \CreaseVec_3 = \rotate{\CreaseVec_1}{\foldAngle_1 / 2} \rotate{\CreaseVec_2}{\foldAngle_2} \CreaseVec_3, \\
	\creaseVec_4 &= \F_3 \CreaseVec_4 = \rotate{\CreaseVec_1}{\foldAngle_1 / 2} \rotate{\CreaseVec_2}{\foldAngle_2} \rotate{\CreaseVec_3}{\foldAngle_3} \CreaseVec_4.
\end{split}
\end{equation}
Given these crease vectors, the following yields normal vectors to $\facet{3}$ and $\facet{4}$, and eventually $\foldAngle_4$:
\begin{equation}
	\begin{split}
		\unNormalVec_{3} &= \creaseVec_3 \times \creaseVec_4, \\
		\foldAngle_4 &= -\sgn\left(\unNormalVec_3 \cdot \etwo\right) \arccos \left(\frac{\unNormalVec_3 \cdot \left(\refl{\etwo} \unNormalVec_3\right)}{\unNormalVec_3 \cdot \unNormalVec_3}\right).
	\end{split}
\end{equation}
An explicit (albeit complicated) expression for $\unNormalVec_3$ is given by
\begin{equation}
\begin{split}
	\unNormalVec_3 = \cSin \foldAngle_1 \cCos \foldAngle_3 \Bigg\{&
		\left[\cSin \foldAngle_1 \left(\cSin \foldAngle_2 + \cTan \foldAngle_3 \left(2 \cCos^2 \sectAngle_1 - \cCos\left(2 \sectAngle_1\right) \cCos \foldAngle_2\right) \right)\right] \eone\\
		&+ \bigg[\frac{\cCos \foldAngle_2}{2} \left(\cCos \sectAngle_1 \cCos \frac{\foldAngle_1}{2} \cTan \foldAngle_3 - 2 \cSin \frac{\foldAngle_1}{2}\right)
			- \cCos 2\sectAngle_1 \cSin \frac{\foldAngle_1}{2} \cSin \foldAngle_2 \cTan \foldAngle_3 \\
			&\qquad - \cCos \sectAngle_1 \cCos \frac{\foldAngle_1}{2} \left(\cSin \foldAngle_2 - 2 \cSin^2 \sectAngle_1 \cTan \foldAngle_3 \right) \bigg] \etwo\\
		&+ \bigg[
		\cCos \frac{\foldAngle_1}{2} \left(\cCos \foldAngle_2 + \cCos \sectAngle_1 \cSin \foldAngle_2 \cTan \foldAngle_3\right) \\
		&\qquad + \frac{\cSin \foldAngle_2 / 2}{2} \left( \cTan\foldAngle_3 \left(\cCos \sectAngle_1 \cCos \foldAngle_2 + 4 \cCos \sectAngle_1 \cSin^2 \sectAngle_1\right) - 2 \cCos \sectAngle_1 \cSin \foldAngle_2\right)
		\bigg] \ethree \Bigg\}.
\end{split}
\end{equation}
The contours of $\foldAngle_4 = \foldAngle_4\left(\foldAngle_2, \foldAngle_3\right)$ are shown in \fref{fig:6-dofs-and-contours_Phi4}.b.
Again, as before with $\foldAngle_1 = \foldAngle_1\left(\foldAngle_2, \foldAngle_3\right)$ (\fref{fig:6-dofs-and-contours}.b), we see that \begin{inparaenum}[1)] \item the cases where $\sectAngle_1 < \pi / 3$ (top row) are characteristically different than the cases where $\sectAngle_1 \geq \pi / 3$ (bottom row) and \item for the case of $\sectAngle_1 = 5 \pi / 12$ (bottom-right), creases $2, 3, 5,$ and $6$ behave collectively as a kind of hinge that requires minimal folding at crease $4$ for a significant range its of motion. \end{inparaenum}
Interestingly, in comparing the bottom-right panels of \fref{fig:6-dofs-and-contours}.b and \fref{fig:6-dofs-and-contours_Phi4}.b, which show, for the case of $\sectAngle_1 = 5 \pi / 12$, the contours of $\foldAngle_1$ and $\foldAngle_4$, respectively, we see that $\foldAngle_4 = \foldAngle_1$ for nearly the whole domain. 
It is as if, in the limit of $\sectAngle_1 \rightarrow \pi / 2$ where creases $2$ and $3$ collapse onto each other, a property of the symmetric 4-vertex is recovered: its two creases within the plane of symmetry have equal and opposite fold angle.
To highlight this, snapshots of folded configurations of an $\sectAngle_1 = 11 \pi / 24$ vertex are shown in \fref{fig:6-dofs-and-contours_Phi4}.a, and an animation of the folded trajectory marked by stars is provided in the ESI (see Animation \#4).

Before moving on, we note an important symmetry inherent in the solutions for the fold angles obtained thus far.
Consider taking a folded origami and flipping it upside-down; then all of the valley folds appear mountain and all of the mountain folds appear valley.
In other words, classifying the folds as mountain or valley implies a choice of $\ethree$.
Thus, there is a symmetry associated with simultaneously making the transformations: $\ethree \rightarrow -\ethree$ and $\text{``valley''} \leftrightarrow \text{``mountain''}$ (and, consequently, $\foldAngle_i \rightarrow -\foldAngle_i, i = 1, \dots, \NCreases$).
We should expect an invariance with respect to this in general.
We give it the name \emph{flip invariance}.
Upon close inspection of \eqref{eq:phi1-6-fold}, one can see that it is indeed flip invariant.
However, flip invariance is also readily illustrated in the contours of $\foldAngle_1 = \foldAngle_1\left(\foldAngle_2, \foldAngle_3\right)$ (\fref{fig:6-dofs-and-contours}) and $\foldAngle_4 = \foldAngle_4\left(\foldAngle_2, \foldAngle_3\right)$ (\fref{fig:6-dofs-and-contours_Phi4}).
Take any point in the $\left(\foldAngle_2, \foldAngle_3\right)$ domain and invert it about the origin, $\left(\foldAngle_2, \foldAngle_3\right) \rightarrow \left(-\foldAngle_2, -\foldAngle_3\right)$.
The contour you arrive at is of the same magnitude, but opposite sign, as expected.

\subsection{$8$-vertex}

Next we consider the crease pattern for a classical origami base: the symmetric $8$-fold waterbomb.
The solution for its fully symmetric, single-degree of freedom folding with alternating mountain and valley creases is well known~\cite{hanna2014waterbomb,hanna2015force,grasinger2022multistability}.
This, however, represents a $1$-dimensional trajectory through its $5$-dimensional configuration space.
Here we use the reflection symmetry to dimensionally reduce the problem to $3$ degrees of freedoms; then solve, simplify, and verify using a computer algebra system.
This allows us to explore a higher dimensional submanifold than in past work.

\begin{figure}
	\centering
	\includegraphics[width=\linewidth]{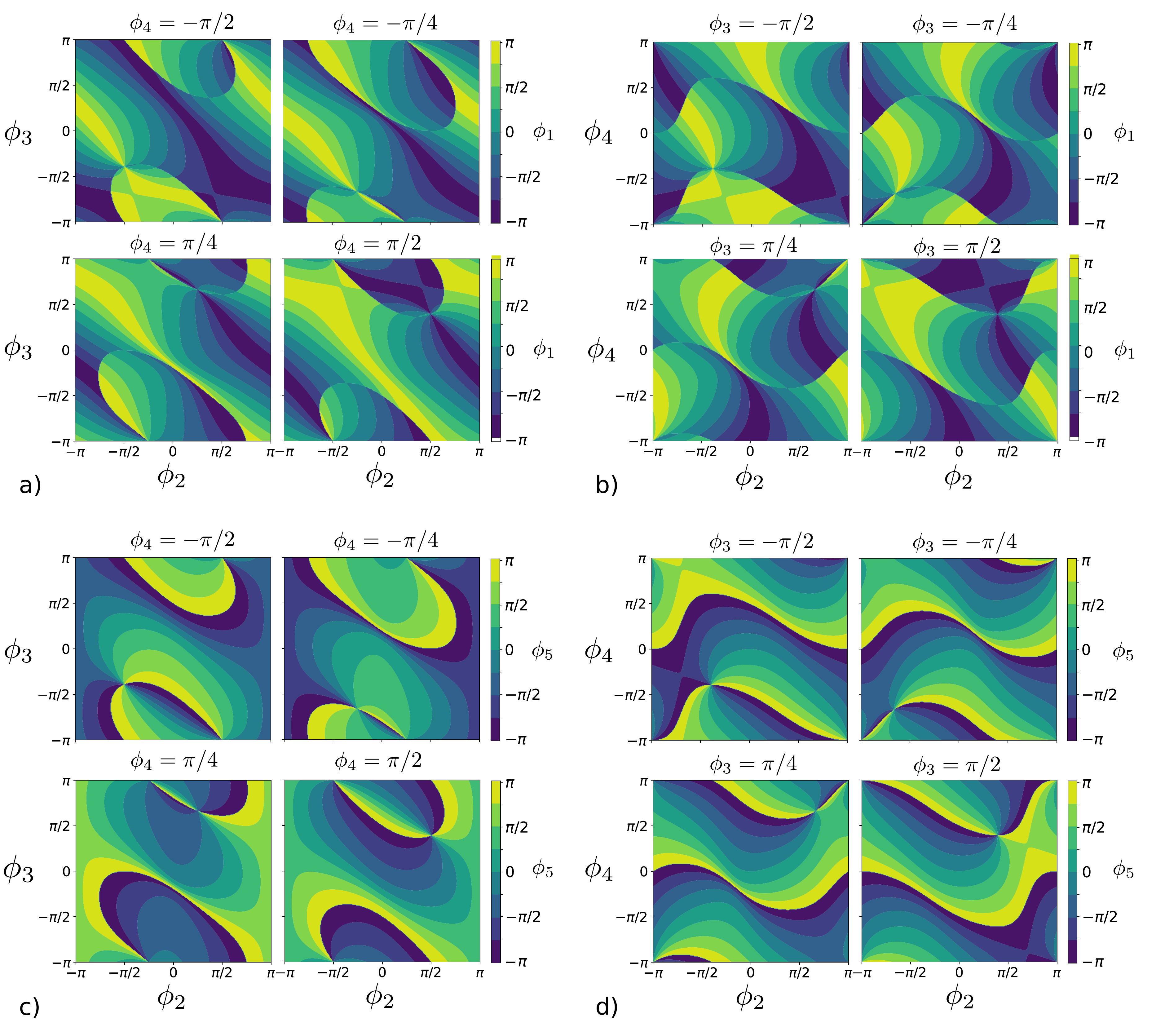}
	\caption{
		\textbf{Kinematics of the degree $8$ vertex with reflection symmetry.}
		Kinematics of a uniform 8-vertex such that the deformed vertices exhibit a reflection symmetry. 
		Contours of slices of $\foldAngle_1 = \foldAngle_1\left(\foldAngle_2, \foldAngle_3, \foldAngle_4\right)$ (top-row) and $\foldAngle_5 = \foldAngle_5\left(\foldAngle_2, \foldAngle_3, \foldAngle_4\right)$ (bottom-row) are shown for fixed $\foldAngle_4 = -\pi / 2, -\pi / 4, \pi / 4,$ and $\pi / 2$ (left column), and fixed $\foldAngle_3 = -\pi / 2, -\pi / 4, \pi / 4,$ and $\pi / 2$ (right column). Note: self-contact not enforced in this visualization.
	}
	\label{fig:8-contours-reflection}
\end{figure}

\newcommand{\auxA}{\mathcal{A}_8}
\newcommand{\auxB}{\mathcal{B}_8}
\newcommand{\auxC}{\mathcal{C}_8}
For a given $\left(\foldAngle_2, \foldAngle_3, \foldAngle_4\right)$, an explicit $3$ degree of freedom solution is given by one of the four solutions\footnote{Analytic determination of which of the four solutions is valid for some $\left(\foldAngle_2, \foldAngle_3, \foldAngle_4\right)$ is difficult, in general.
However, when working numerically, it is trivial to plug $\foldAngle_1, \dots, \foldAngle_8$ into \eqref{eq:compatibility} for each of the four solutions and check which results in the least error.}
\begin{subequations} \label{eq:8-reflection}
\begin{equation} \label{eq:8-reflection-1}
	\foldAngle_1 = \begin{cases}
		\convToFoldAngle\left(2 \arctan\left(\argSign_8 \left(\auxA + \auxB\right), \argSign_8 \auxC\right)\right) \\
		\convToFoldAngle\left(2 \arctan\left(\argSign_8 \auxC\right), \argSign_8 \left(\auxA - \auxB\right)\right)
	\end{cases},
\end{equation}
where $\argSign_8 = \pm 1$, and where
\begin{equation} \label{eq:8-reflection-2}
\begin{split}
	\auxA\left(\foldAngle_2, \foldAngle_3, \foldAngle_4\right) \coloneqq & \: 4 \cos \left(\foldAngle_2 + \frac{\foldAngle_4}{2}\right) \cos \frac{\foldAngle_4}{2} \sin \foldAngle_3, \\
	\auxB\left(\foldAngle_2, \foldAngle_3, \foldAngle_4\right) \coloneqq & \: \sqrt{2} \left( \sin \foldAngle_2 \left(1 + \cos\foldAngle_3 + \left(\cos\foldAngle_3 - 1\right) \cos \foldAngle_4\right) + 2 \cos\foldAngle_2 \cos\foldAngle_3 \sin\foldAngle_4 \right), \\
	\auxC\left(\foldAngle_2, \foldAngle_3, \foldAngle_4\right) \coloneqq & \: 1 - \cos\foldAngle_4 - \sqrt{2} \sin\foldAngle_2 \sin\foldAngle_3 \left(1 + \cos\foldAngle_4\right) + \sqrt{2} \sin\foldAngle_3 \sin\foldAngle_4 -\\
	& \cos\foldAngle_3 \left(1 + \cos\foldAngle_4 + 2 \sin\foldAngle_2 \sin\foldAngle_4\right) +\\
	& \cos\foldAngle_2\left(1 + \cos\foldAngle_3 + \cos\foldAngle_4 \left(\cos\foldAngle_3 - 1\right) - \sqrt{2} \sin\foldAngle_3 \sin\foldAngle_4\right).
\end{split}
\end{equation}
\end{subequations}
The minimum numerical error was tested extensively and found, that in every case considered, the error was within numerical precision (i.e. $\lVert \prod_{i=1}^8 \rotAbout{\CreaseVec_i}\left(\foldAngle_i\right) - \Iden \rVert_2 \lessapprox 10^{-16}$).
Given $\foldAngle_1, \dots, \foldAngle_4$, the fold angle $\foldAngle_5$ was computed by
\begin{equation}
\begin{split}
\unNormalVec_{4} &= \creaseVec_4 \times \creaseVec_5, \\
\foldAngle_5 &= -\sgn\left(\unNormalVec_4 \cdot \etwo\right) \arccos \left(\frac{\unNormalVec_4 \cdot \left(\refl{\etwo} \unNormalVec_4\right)}{\unNormalVec_4 \cdot \unNormalVec_4}\right).
\end{split}
\end{equation}
Contours of slices of $\foldAngle_1 = \foldAngle_1\left(\foldAngle_2, \foldAngle_3, \foldAngle_4\right)$ (top-row) and $\foldAngle_5 = \foldAngle_5\left(\foldAngle_2, \foldAngle_3, \foldAngle_4\right)$ (bottom-row) are shown in \fref{fig:8-contours-reflection} for fixed $\foldAngle_4 = -\pi / 2, -\pi / 4, \pi / 4,$ and $\pi / 2$ (left column), and fixed $\foldAngle_3 = -\pi / 2, -\pi / 4, \pi / 4,$ and $\pi / 2$ (right column).
The information rich equation \eqref{eq:8-reflection} and \fref{fig:8-contours-reflection} show a reduced picture of the many different possible ways of folding, and associated degrees of freedom, for the symmetric $8$-fold waterbomb vertex; however, the expressions are complex.
In the next section, we extend the approach to allow for further reduction of dimensionality and exploitation of symmetry.

\section{Vertices with combined rotations and reflections} \label{sec:vertices-with-reflection-and-rotation}

\subsection{$8$-vertex} \label{sec:8-rotation-reflection}

Next we adjust our choice (up to rigid body rotations) of the deformation map in order to exploit symmetry groups which also include rotations.
Inspired by the kinematic parameterization of the fully symmetric folding for the $8$-fold waterbomb used in Hanna et al.~\cite{hanna2014waterbomb}, we consider first elevating crease $1$ by rotating via $\rotate{\etwo}{\elevAngle}$\footnote{In~\cite{hanna2014waterbomb,hanna2015force,grasinger2022multistability}, the analogous parameter is the angle, $\theta$, between the vertical axis (i.e. unit direction orthogonal to the flat state) and the crease vector for the valley creases.} and then deforming in the usual way, i.e.
\begin{equation} \label{eq:fold-map-rotation-reflection}
	\F_i = \begin{cases}
		\rotate{\etwo}{\elevAngle} \rotate{\CreaseVec_1}{\partFoldAngle_1}, & i = 1, \\
		\rotate{\etwo}{\elevAngle} \rotate{\CreaseVec_1}{\partFoldAngle_1} \prod_{j=2}^{i} \rotate{\CreaseVec_j}{\foldAngle_j}, & i \neq 1
	\end{cases}.
\end{equation}
where we introduce notation $\partFoldAngle_1$ to denote the right \emph{partial fold angle} at crease $1$ (i.e. the contribution to the fold angle on the $+\etwo$ side of the crease).
Previously, it was the case that $\partFoldAngle_1 = \foldAngle_1 / 2$, but for vertices with rotational symmetry this relationship can be relaxed and compatible states with less symmetry (such that the left and right partial fold angles are not necessarily equal) can be generated.

\begin{figure}
	\centering
	\includegraphics[width=0.85\linewidth]{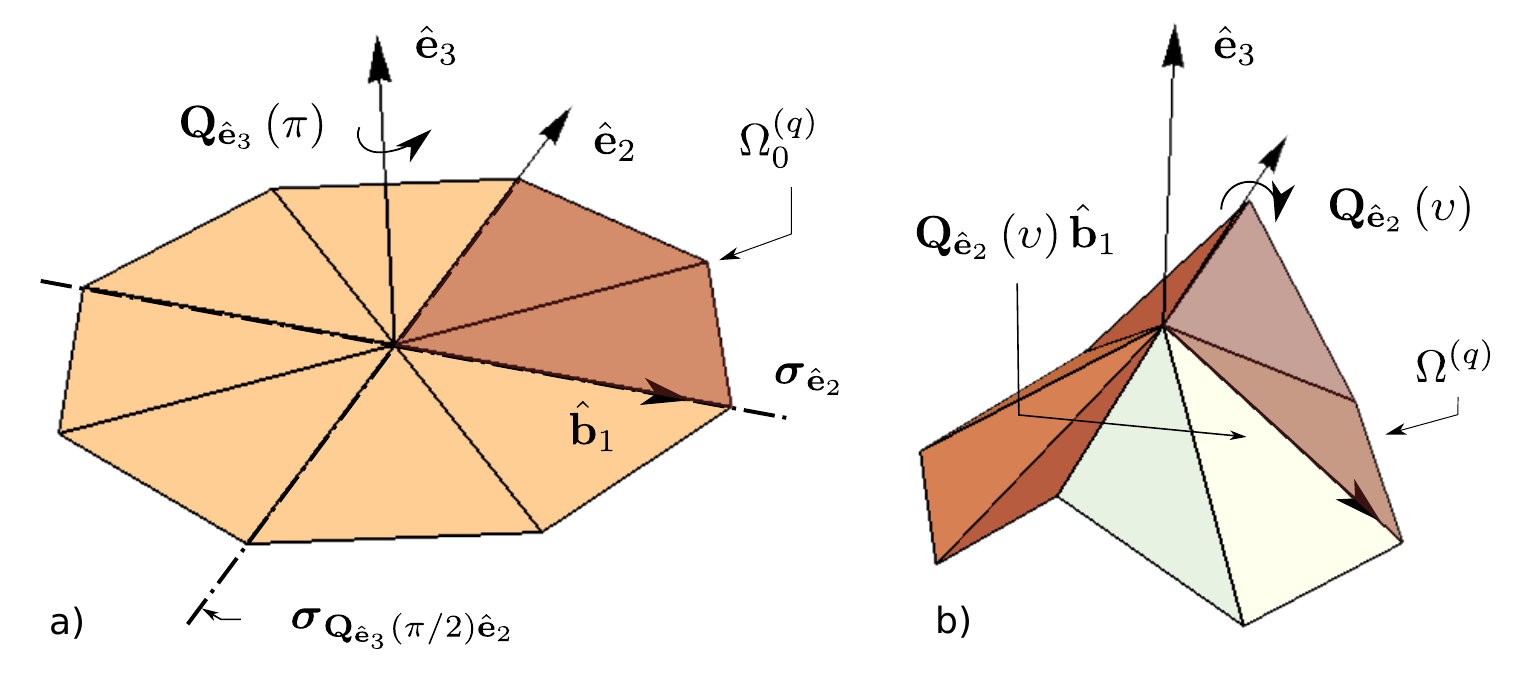}
	\caption{
		\textbf{Deformation map for vertices with reflection and rotational symmetries.}
		a) The flat, reference configuration of a uniform degree $8$ vertex can be described by $\orbit{\Facet{q}}{\group}$ where $\Facet{q} = \Facet{1} \bigcup \Facet{2}$ and $\group = \genBy{\refl{\etwo}, \rotate{\ethree}{\pi}}$.
		b) The deformation map is chosen, up to rigid body rotations, such that $\Facet{q}$ is first rotated by some angle of elevation $\elevAngle$ and then folded; the modified parameterization of folded states allows the symmetry of the folded state to also be described by $\group$ as $\ethree$ can then remain the axis of the rotational symmetry.
	}
	\label{fig:rotation-reflection-deformation-map}
\end{figure}

The parameterization introduced in \eqref{eq:fold-map-rotation-reflection} allows us to describe the symmetry of the folded state using the same symmetry group as the flat state.
Indeed, let $\group = \genBy{\refl{\etwo}, \rotate{\ethree}{\pi}}$ where $\genBy{\generic}$ is the group generated by $\generic$; and let $\Facet{q} = \Facet{1} \bigcup \Facet{2}$.
Then \fref{fig:rotation-reflection-deformation-map}.a shows that the flat state of the $8$-fold waterbomb satisfies the symmetry described by $\group$; and \fref{fig:rotation-reflection-deformation-map}.b shows an example folded configuration whose deformation is described by \eqref{eq:fold-map-rotation-reflection} and also satisfies the symmetry described by $\group$.

Geometrically, the significance of this reparameterization for the deformation map consists in its ability to use the orbit of the same group to construct the folded configuration from $\facet{q}$ as the group used to construct the flat configuration from $\Facet{q}$.
For the purposes of generalization, we now explore this reparamterization in its algebraic context.
Note that, by \eqref{eq:two-reflections-is-a-rotation} and \eqref{eq:perpendicular-reflections-commute}, $\rotate{\ethree}{\pi} = \refl{\etwo}\refl{\eone} = \refl{\eone}\refl{\etwo}$.
Assume for now that $\partFoldAngle_1 = \foldAngle_2 / 2$.
Beginning with \eqref{eq:compatibility}:
\begin{equation}
\begin{split}
	\rotate{\etwo}{\elevAngle} 
	\rotate{\CreaseVec_1}{\partFoldAngle_1} 
	\rotate{\CreaseVec_2}{\foldAngle_2} 
	\rotate{\CreaseVec_3}{\foldAngle_3} 
	\rotate{\refl{\eone} \CreaseVec_2}{\foldAngle_4}
	\rotate{\refl{\eone} \CreaseVec_1}{\foldAngle_5} \times& \\
	\quad \rotate{\rotate{\ethree}{\pi}\CreaseVec_2}{\foldAngle_6} 
	\rotate{\refl{\etwo} \CreaseVec_3}{\foldAngle_7} 
	\rotate{\refl{\etwo} \CreaseVec_2}{\foldAngle_8} 
	\rotate{\refl{\etwo} \CreaseVec_1}{\partFoldAngle_1} 
	\rotate{\etwo}{-\elevAngle} &= \Iden.
\end{split}
\end{equation}
Then by \eqref{eq:identities}
\begin{equation} \label{eq:8-vertex-compat-1}
\begin{split}
\rotate{\etwo}{\elevAngle} 
\rotate{\CreaseVec_1}{\partFoldAngle_1} 
\rotate{\CreaseVec_2}{\foldAngle_2} 
\rotate{\CreaseVec_3}{\foldAngle_3} 
\refl{\eone} 
\rotate{\CreaseVec_2}{-\foldAngle_4} 
\rotate{\CreaseVec_1}{-\foldAngle_5} \times& \\
\quad \refl{\etwo} 
\rotate{\CreaseVec_2}{\foldAngle_6}
\refl{\eone}
\rotate{\CreaseVec_3}{-\foldAngle_7} 
\rotate{\CreaseVec_2}{-\foldAngle_8}
\rotate{\CreaseVec_1}{\partFoldAngle_1} 
\rotate{\etwo}{-\elevAngle}
\refl{\etwo} &= \Iden.
\end{split}
\end{equation}

Now we return to the proposed ansatz for generating a compatible folded configuration.
For instance, if $\facet{3} = \refl{\eone} \facet{2} \subset \orbit{\facet{q}}{\group}$ as proposed, then this implies that
\begin{equation} \label{eq:rotation-reflection-8-full}
	\rotate{\etwo}{\elevAngle} 
	\rotate{\CreaseVec_1}{\partFoldAngle_1} 
	\rotate{\CreaseVec_2}{\foldAngle_2} 
	\rotate{\CreaseVec_3}{\foldAngle_3} = \refl{\eone} \rotate{\etwo}{\elevAngle} 
	\rotate{\CreaseVec_1}{-\partFoldAngle_1} 
	\rotate{\CreaseVec_2}{\foldAngle_2} \refl{\eone}.
\end{equation}
Assume $\partFoldAngle_1, \foldAngle_2,$ and $\foldAngle_3$ satisfies \eqref{eq:rotation-reflection-8-full}; further, let $\foldAngle_4 = \foldAngle_2$ and $\foldAngle_5 = \foldAngle_1 = 2 \partFoldAngle_1$.
Then, plugging into \eqref{eq:8-vertex-compat-1}, we obtain
\begin{equation}
	\refl{\eone} \rotate{\etwo}{\elevAngle} 
	\rotate{\CreaseVec_1}{-\partFoldAngle_1}
	\refl{\etwo} 
	\rotate{\CreaseVec_2}{\foldAngle_6}
	\refl{\eone}
	\rotate{\CreaseVec_3}{-\foldAngle_7} 
	\rotate{\CreaseVec_2}{-\foldAngle_8}
	\rotate{\CreaseVec_1}{-\partFoldAngle_1} 
	\rotate{\etwo}{-\elevAngle} 
	\refl{\etwo} = \Iden.
\end{equation}
Next, let $\foldAngle_6 = \foldAngle_2, \foldAngle_7 = \foldAngle_3,$ and $\foldAngle_8 = \foldAngle_2$ (since $\CreaseVec_6 = \rotate{\ethree}{\pi} \CreaseVec_2, \CreaseVec_7 = \refl{\etwo} \CreaseVec_3,$ and $\CreaseVec_8 = \refl{\etwo} \CreaseVec_2$, respectively).
Then using \eqref{eq:rotation-reflection-8-full} again
\begin{equation} \label{eq:8-vertex-compat-2}
\refl{\eone} \rotate{\etwo}{\elevAngle} 
\rotate{\CreaseVec_1}{-\partFoldAngle_1}
\refl{\etwo}
\rotate{\CreaseVec_1}{-\partFoldAngle_1} 
\rotate{\etwo}{-\elevAngle}
\refl{\eone}
\refl{\etwo} = \Iden.
\end{equation}
Recalling $\rotate{\CreaseVec_1}{-\partFoldAngle_1} = \refl{\etwo} \rotate{\CreaseVec_1}{\partFoldAngle_1} \refl{\etwo}$, $\rotate{\etwo}{\elevAngle} \refl{\etwo} = \refl{\etwo} \rotate{\etwo}{\elevAngle}$, and $\refl{\eone} \refl{\etwo} = \refl{\etwo} \refl{\eone}$, we see that equality is indeed obtained in \eqref{eq:8-vertex-compat-2}, as desired.
From this we can conclude that \eqref{eq:rotation-reflection-8-full} represents a reduced-order, $2$ degree of freedom model for the symmetric $8$-fold waterbomb.
As in previous examples, we can reformulate \eqref{eq:rotation-reflection-8-full} as an equivalent but simplified form; indeed,
\begin{equation} \label{eq:rotation-reflection-8}
	\left(\rotate{\etwo}{\elevAngle} 
	\rotate{\CreaseVec_1}{\partFoldAngle_1} 
	\rotate{\CreaseVec_2}{\foldAngle_2} 
	\CreaseVec_3\right) \cdot \eone = 0,
\end{equation}
is an equivalent condition which can readily be used to solve for $\foldAngle_1 = \foldAngle_1\left(\elevAngle, \foldAngle_2\right) = 2 \partFoldAngle_1\left(\elevAngle, \foldAngle_2\right)$.
Then one can solve for the remaining fold angle, $\foldAngle_3$, by computing $\creaseVec_2 = \F_1 \CreaseVec_2$, $\creaseVec_3 = \F_2 \CreaseVec_3$, and subsequently, a vector normal to $\facet{2}$: $\unNormalVec_2 = \creaseVec_2 \times \creaseVec_3$.
Lastly,
\begin{equation}
	\foldAngle_3 = -\sgn\left(\unNormalVec_2 \cdot \eone\right) \arccos \left(\frac{\unNormalVec_2 \cdot \refl{\eone} \unNormalVec_2}{\left|\unNormalVec_2\right|^2}\right).
\end{equation}

To summarize, \begin{inparaenum}[1)] \item we choose $\elevAngle$ and $\foldAngle_2$, \item solve \eqref{eq:rotation-reflection-8} for $\partFoldAngle_1$, and finally \item make the assignments
\begin{equation}
\foldAngle_4 = \foldAngle_6 = \foldAngle_8 = \foldAngle_2, \quad \foldAngle_5 = \foldAngle_1 = 2 \partFoldAngle_1, \quad \foldAngle_7 = \foldAngle_3.
\end{equation}
\end{inparaenum}

\begin{figure}
	\centering
	\includegraphics[width=\linewidth]{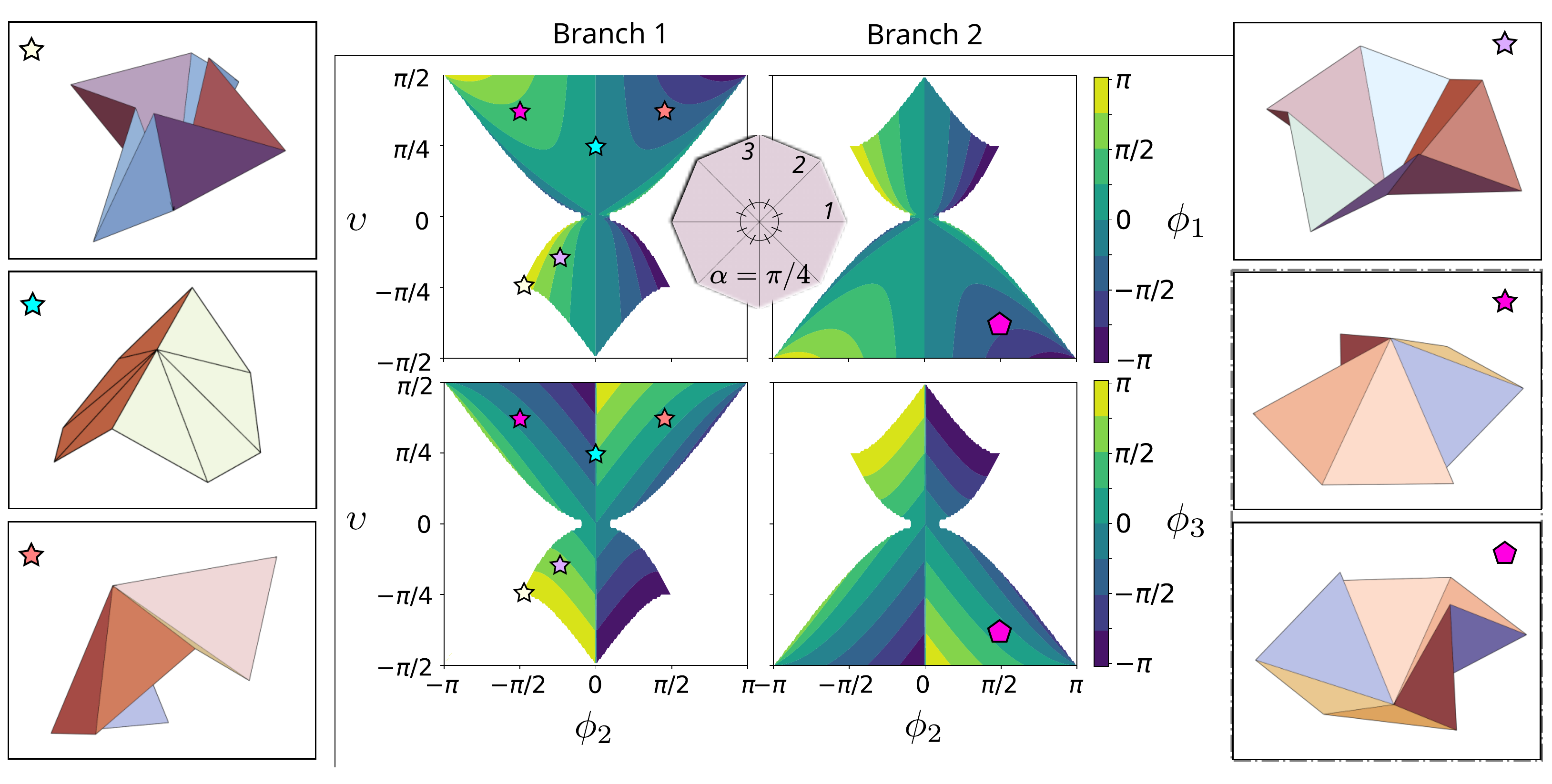}
	\caption{
		\textbf{Degrees of freedom and kinematics with $\dihedral{2}$ symmetry of a uniform degree $8$ vertex.}
		Contours of $\foldAngle_1 = \foldAngle_1\left(\elevAngle, \foldAngle_2\right)$ (top row) and $\foldAngle_3 = \foldAngle_3\left(\elevAngle, \foldAngle_2\right)$ (bottom row), and example folded states of a uniform degree $8$ vertex such that the folded states satisfy the symmetry corresponding to the group $\genBy{\refl{\etwo}, \rotate{\ethree}{\pi}}$.
		White regions of the contour plots signify parts of the $\left(\elevAngle, \foldAngle_2\right)$ domain which are inadmissible; that is, a folded configuration with the specified $\left(\elevAngle, \foldAngle_2\right)$ and symmetry cannot be attained.
		The admissible space consists of two distinct regions connected only at the flat state (i.e. the origin).
		Physically, for there to be any folding about any of the creases without stretching the facets themselves, it is required that the center vertex flex in some way.
		When a point is inverted about the origin and moved to the other folding solution branch, $\foldAngle_1$ and $\foldAngle_3$ have the same magnitude but opposite sign, which means the solutions satisfy flip invariance.
	}
	\label{fig:8-dofs-and-contours-rotation-reflection}
\end{figure}

\begin{figure}
	\centering
	\includegraphics[width=\linewidth]{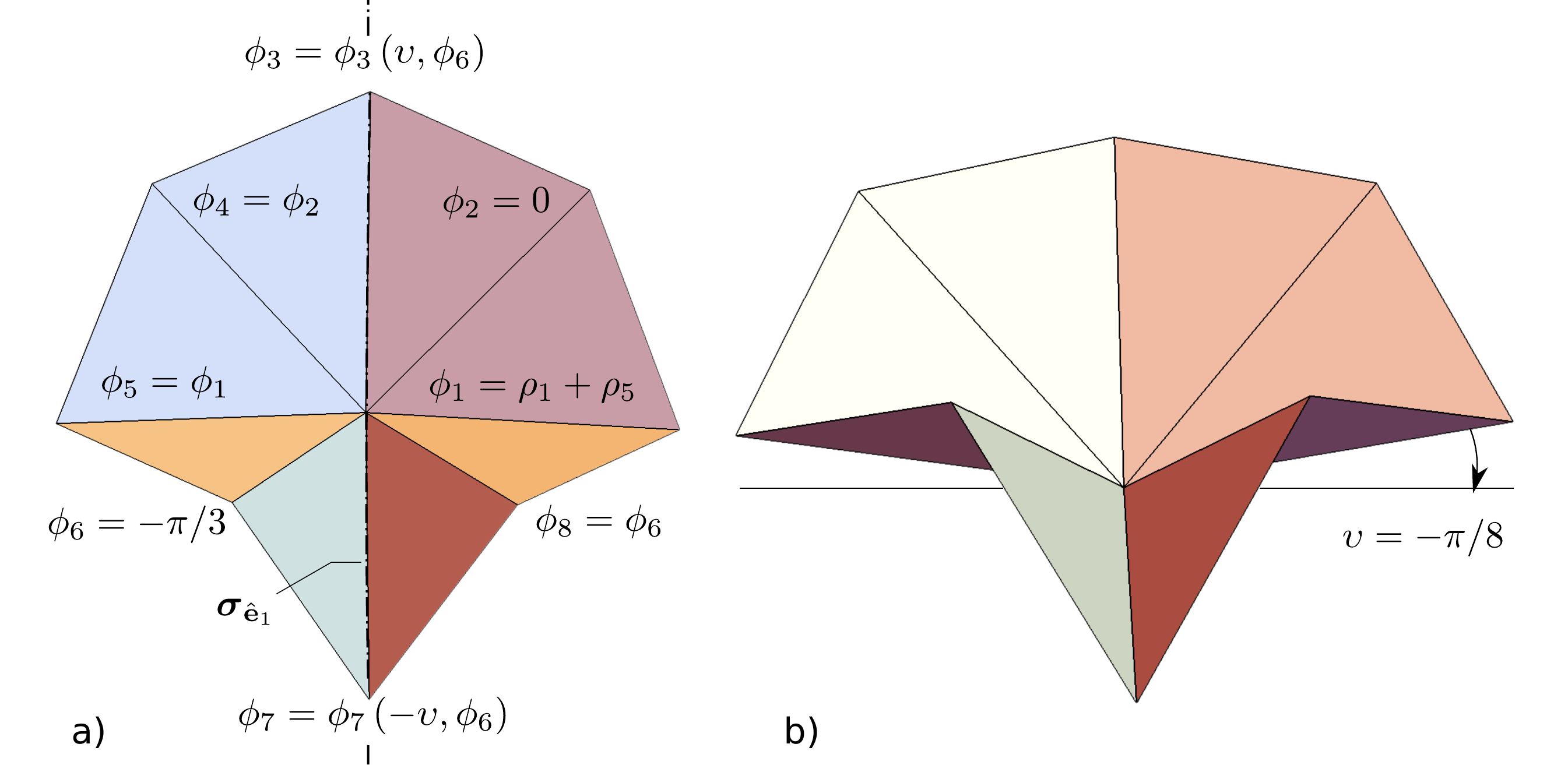}
	\caption{
		\textbf{Constructing compatibly folded states with symmetry breaking.}
		a) Leveraging the solution of the reduced order model with two degrees of freedom (i.e. $\elevAngle$ and $\foldAngle_2$) to generate a solution with less symmetry and three degrees of freedom.
		In this example, $\elevAngle = -\pi / 8$, $\foldAngle_2 = 0$, and $\foldAngle_6 = -\pi / 3 \neq \foldAngle_2$.
		b) Alternate view.
	}
	\label{fig:8-folded-mixed}
\end{figure}

\newcommand{\auxD}{\mathcal{D}}
\newcommand{\auxF}{\mathcal{F}}
The solution to \eqref{eq:rotation-reflection-8} is given by
\begin{equation} \label{eq:8-phi1-rr}
	\partFoldAngle_1 = \begin{cases}
		\begin{cases}
		\arctan\left( \quad
			\sqrt{2} \auxF \left(\cos\foldAngle_2-1\right) - \auxD\left(\cos\foldAngle_2+1\right), 
			\quad \sin\foldAngle_2\left(\auxD\sqrt{2} - \auxF\right)
		\, \right) & \foldAngle_2 > 0 \\
		\arctan\left( \:
			-\sqrt{2} \auxF \left(\cos\foldAngle_2-1\right) - \auxD\left(\cos\foldAngle_2+1\right), 
			-\sin\foldAngle_2\left(\auxD\sqrt{2} + \auxF\right)
		\, \right) & \foldAngle_2 < 0
		\end{cases} \\
		\begin{cases}
		\arctan\left( \quad
			\sqrt{2} \auxF \left(\cos\foldAngle_2-1\right) + \auxD\left(\cos\foldAngle_2+1\right), 
			\quad \sin\foldAngle_2\left(\auxD\sqrt{2} + \auxF\right)
		\, \right) & \foldAngle_2 > 0 \\
		\arctan\left( \:
			-\sqrt{2} \auxF \left(\cos\foldAngle_2-1\right) + \auxD\left(\cos\foldAngle_2+1\right), 
			-\sin\foldAngle_2\left(\auxD\sqrt{2} - \auxF\right)
		\, \right) & \foldAngle_2 < 0
		\end{cases}
	\end{cases}
\end{equation}
where there are two solution branches given by the pair of equations delimited by the top inner braces and the pair of equations delimited by the bottom inner braces; and where
\begin{equation}
\begin{split}
	\auxD\left(\elevAngle, \foldAngle_2\right) &\coloneqq \sqrt{\sin^2 \foldAngle_2 + 2 \cos \foldAngle_2 - 2 \cos\left(2 \elevAngle\right)}, \\
	\auxF\left(\elevAngle, \foldAngle_2\right) &\coloneqq \left|\sin\elevAngle \sin\foldAngle_2\right| \cot\elevAngle.
\end{split}
\end{equation}
Consequently, the solution is only real valued (and therefore valid) provided: $\sin^2 \foldAngle_2 + 2 \cos \foldAngle_2 - 2 \cos\left(2 \elevAngle\right) \geq 0$.
Let $\foldAngle_1 = \convToFoldAngle\left(2 \partFoldAngle\right)$. 
Contours of $\foldAngle_1 = \foldAngle_1\left(\elevAngle, \foldAngle_2\right)$ (top row) and $\foldAngle_3 = \foldAngle_3\left(\elevAngle, \foldAngle_2\right)$ (bottom row), for both solutions (folding branch 1 and 2 are the left and right columns, respectively) are shown in \fref{fig:8-dofs-and-contours-rotation-reflection}, as well as example deformed configurations.
There are multiple symmetries and other properties of \eqref{eq:8-phi1-rr} and \fref{fig:8-dofs-and-contours-rotation-reflection} that should be noted.
First, notice that \eqref{eq:8-phi1-rr} and \fref{fig:8-dofs-and-contours-rotation-reflection} exhibit flip invariance, as expected, where the ``flipping'' operation now also includes $\elevAngle \rightarrow -\elevAngle$, and a change of folding solution branch.
Flip invariance is illustrated in \fref{fig:8-dofs-and-contours-rotation-reflection} by observing that $\left(\elevAngle, \foldAngle_2\right) \rightarrow \left(-\elevAngle, -\foldAngle_2\right)$ results in a contour of the same magnitude but opposite sign.
In the bottom right panel, delimited by a dashed rectangle, we show two folded states which are related by inversion through the origin and change of folding solution branch. 
Physically, this operation corresponds with flipping the origami over (i.e. reflection via $\refl{\ethree}$).
Second, while there are multiple solutions for $\foldAngle_1 = \foldAngle_1\left(\elevAngle, \foldAngle_2\right)$ (top row), one can be obtained from the other by the change of variables $\foldAngle_2 \rightarrow -\foldAngle_2$; i.e. the contours in branch 2 (right column) are simply the contours of branch 1 (left column) reflected about $\foldAngle_2 = 0$.
The contours of $\foldAngle_3 = \foldAngle_3\left(\elevAngle, \foldAngle_2\right)$ (bottom row) have a similar property: we can obtain the branch 2 from branch 1 by reflecting about $\elevAngle = 0$.
Lastly, the condition $\foldAngle_1 \in \Reals$ (equivalently: $\sin^2 \foldAngle_2 + 2 \cos \foldAngle_2 - 2 \cos\left(2 \elevAngle\right) \geq 0$) results in two separate regions for each of the contour plots shown in \fref{fig:8-dofs-and-contours-rotation-reflection}: one where the center vertex of the origami is up (i.e. $\elevAngle > 0$) and one where it is down (i.e. $\elevAngle < 0$).
The two separate regions only meet at a single point where $\foldAngle_1 = \foldAngle_2 = \dots = \foldAngle_8 = \elevAngle = 0$, or i.e. the flat configuration.
This can be understood as a consequence of the rigidity of the facets.
For the vertex to fold with the symmetry of interest and without stretching the facets themselves, it is required that the center vertex flex in some way and, consequently, $\elevAngle \neq 0$.
As alluded to previously, there is a subtle and interesting property of the solutions generated by \eqref{eq:rotation-reflection-8}.
Because of the symmetry of the underlying vertex and the convention chosen for the deformation map, we have, respectively, that $\CreaseVec_{5} = \rotate{\ethree}{\pi} \CreaseVec_1$ and $\creaseVec_{5} = \rotate{\ethree}{\pi} \creaseVec_1$; together, these imply that $\creaseVec_5 = \rotate{\etwo}{-\elevAngle} \CreaseVec_5$.
But this was precisely the starting point for constructing solutions for the first half of the vertex with the exception that crease $5$ has now taken the place of crease $1$ and $\elevAngle \rightarrow -\elevAngle$.
Let the condition $\cBody = \orbit{\facet{q}}{\group}$ be relaxed.
Instead, consider the subgroup $\otherGroup = \set{\Iden, \refl{\eone}}$.
Recognizing that a compatible folded state of the remaining subset of the vertex can be described by the same reduced order model where crease $5$ is analogous to crease $1$ and creases $6$ through $8$ are analogous to creases $2$ through $4$, we adjust our convention for the deformation map and make the following definitions:
\begin{equation} \label{eq:fold-map-symmetry-breaking}
	\F_i = \begin{cases}
	\rotate{\etwo}{\elevAngle} \rotate{\CreaseVec_1}{\partFoldAngle_1}, & i = 1, \\
	\rotate{\etwo}{\elevAngle} \rotate{\CreaseVec_1}{\partFoldAngle_1} \prod_{j=2}^{i} \rotate{\CreaseVec_j}{\foldAngle_j}, & 1 < i \leq 4 \\
	\rotate{\etwo}{-\elevAngle} \rotate{\CreaseVec_5}{\partFoldAngle_5} \qquad \left(= \F_4 \rotate{\CreaseVec_5}{\partFoldAngle_5}\right), & i = 5, \\
	\rotate{\etwo}{-\elevAngle} \rotate{\CreaseVec_5}{\partFoldAngle_5} \prod_{j=6}^{i} \rotate{\CreaseVec_j}{\foldAngle_j} \quad \left(= \F_4 \rotate{\CreaseVec_5}{\partFoldAngle_5} \prod_{j=6}^{i} \rotate{\CreaseVec_j}{\foldAngle_j}\right), & \text{otherwise}
	\end{cases},
\end{equation}
and
\begin{equation}
\begin{split}
	\Facet{r} &\coloneqq \Facet{5} \bigcup \Facet{6}, \quad \facet{r} \coloneqq \facet{5} \bigcup \facet{6} \\
	\Facet{\ell} &\coloneqq \orbit{\Facet{r}}{\group}, \quad \facet{\ell} \coloneqq \orbit{\facet{r}}{\group}
\end{split}
\end{equation}
Given \eqref{eq:fold-map-symmetry-breaking}, we can generate a compatibly folded state \emph{with less symmetry than $\dihedral{2}$} by \begin{enumerate}
	\item specifying $\elevAngle$ and $\foldAngle_2$,
	\item solving for $\partFoldAngle_1$ and $\foldAngle_3$ using \eqref{eq:8-phi1-rr} (and the deformation map),
	\item specifying $\foldAngle_6$,
	\item solving for $\partFoldAngle_5$ and $\foldAngle_7$, again using \eqref{eq:8-phi1-rr} with appropriate analogs (e.g. crease $1 \rightarrow$ crease $5$, crease $2 \rightarrow$ crease $6$, etc.),
	and
	\item making the assignments
	\begin{equation}
		\foldAngle_1 = \foldAngle_5 = \partFoldAngle_1 + \partFoldAngle_5, \quad \foldAngle_4 = \foldAngle_2, \quad \foldAngle_8 = \foldAngle_6.
	\end{equation}
\end{enumerate}
Through this procedure is we have leveraged the solution of a reduced order model with two degrees of freedom (i.e. $\elevAngle$ and $\foldAngle_2$) to generate a solution with less symmetry and three degrees of freedom (i.e. $\elevAngle, \foldAngle_2$ and $\foldAngle_6$).
Note that the symmetry is now described by $\otherGroup \subgroup \group$ such that we have broken the symmetry associated with the $\refl{\etwo}$ transformation.
As we will see in future developments, this procedure generalizes to all of the examples that we consider where the folded state has both reflection and rotational symmetries.
It represents an analytical tool for generating families of solutions, generally with less symmetry and often with more degrees of freedom, from a simpler, lower dimensional equation.
An example folded configuration of the symmetric $8$-fold vertex generated by this procedure is shown in \fref{fig:8-folded-mixed}.

\subsection{Generalized reflection-rotations} \label{sec:general-rotation-reflection}

Next we are interested in how the method developed in \fref{sec:8-rotation-reflection} can be generalized to vertices with symmetries described by groups, $\group$, where $\group = \genBy{\refl{\etwo}, \rotate{\ethree}{\genAngle}}$ for general angle $\genAngle$\footnote{
    
     If $\group = \genBy{\refl{\etwo}, \rotate{\ethree}{\genAngle}}$, then this implies that there are also 
     reflection symmetries about planes normal to $\rotAbout{\ethree}^m\left(\genAngle / 2\right) \etwo$ for all $m \in \left\{ j \in \Naturals : j \leq \pi / \genAngle\right\}$. 
    
}.
First, note that for $\group$ to satisfy the closure property, we need $\rotAbout{\ethree}^m\left(\genAngle\right) = \Iden$ for some $m \in \Integers$.
In addition, the choice of deformation map and overall approach used to derive \eqref{eq:rotation-reflection-8} made use of the fact that the two planes of reflection symmetries cut through creases of the vertex (e.g. creases, $1, 3, 5,$ and $7$.).
This can always be done provided $\genAngle = 2 \pi / 2 h$, and, consequently, $\group = \dihedral{2 h}$ where $h \in \Naturals$ and $\dihedral{2 h}$ is the group of symmetries of a regular polygon of $2h$ sides.
To this end, let $\genAngle = 2 \pi / 2 h, h \in \Naturals$, $\CreaseVec_k = \rotate{\ethree}{\genAngle / 2} \CreaseVec_1$ (i.e. $\refl{\CreaseVec_k \times \ethree} \in \group$), and $\Facet{c} = \bigcup_{i=1}^{k-1} \Facet{i}$, where $\Facet{c} \subset \rBody$ is the unit cell to be folded such that $\cBody = \orbit{\facet{c}}{\group}$.
Then
\begin{equation} \label{eq:general-reflection-rotation-consistency-condition}
	\frac{\creaseVec_{k} \cdot \etwo}{\creaseVec_{k} \cdot \eone} = \frac{\left(\F_{k-1}\left(\elevAngle, \foldAngle_1, \dots, \foldAngle_{k-1}\right) \CreaseVec_{k}\right) \cdot \etwo}{\left(\F_{k-1}\left(\elevAngle, \foldAngle_1, \dots, \foldAngle_{k-1}\right) \CreaseVec_k\right) \cdot \eone} = \frac{\CreaseVec_k \cdot \etwo}{\CreaseVec_k \cdot \eone},
\end{equation}
represents a reduced-order model for solutions to \eqref{eq:compatibility} with $\dihedral{2 h}$ symmetry where, when the convention chosen for the deformation map is given by \eqref{eq:fold-map-rotation-reflection}, the axis of rotation for the rotational symmetries is $\ethree$ and the normal vectors for each of the planes of reflection are orthogonal to $\ethree$.
In principle, one \begin{inparaenum}[1)] \item solves \eqref{eq:general-reflection-rotation-consistency-condition} for $\foldAngle_1 = \foldAngle_1\left(\elevAngle, \foldAngle_2, \dots, \foldAngle_{k-1}\right)$, then \item solves for $\foldAngle_k$ by\footnote{
The sign term adheres to the mountain-valley convention by recognizing that if a crease lies within a plane of reflection, it is mountain if the normal of its right facet shares a similar direction with the normal to the plane of symmetry (provided the sign of the normal is chosen as in \eqref{eq:general-reflection-rotation-other-fold-angle}).
}
\begin{equation} \label{eq:general-reflection-rotation-other-fold-angle}
\begin{split}
	\foldAngle_k &= -\sgn\left(\unNormalVec_{k-1} \cdot \left(\CreaseVec_k \times \ethree\right)\right) \arccos \left(\frac{\unNormalVec_{k-1} \cdot \refl{\CreaseVec_k \times \ethree} \unNormalVec_{k-1}}{\left|\unNormalVec_{k-1}\right|^2}\right), \\
	\unNormalVec_{k-1} &= \creaseVec_{k-1} \times \creaseVec_k,
\end{split}
\end{equation}
and finally \item makes the assignments
\begin{equation} \label{eq:general-reflection-rotation-angle-assignments}
	\foldAngle_{k+1} = \foldAngle_{k-1}, \foldAngle_{k+2} = \foldAngle_{k-2}, \cdots, \foldAngle_{2k-1} = \foldAngle_1, \foldAngle_{2 k} = \foldAngle_2, \foldAngle_{2 k + 1} = \foldAngle_3, \dots, \text{etc.};
\end{equation}
\end{inparaenum}
in other words, after crease $k$, going counterclockwise around the vertex, the fold angles are assigned in decreasing order until $1$, then increasing order until $k$, then decreasing until $1$, ..., etc., until all of the fold angles have been assigned.
For the example considered in \fref{sec:8-rotation-reflection}, we had $k = 3, \: \genAngle = \pi$ and the sequence of fold angle assignments went $\left(\foldAngle_1, \dots, \foldAngle_8\right) = \left(\foldAngle_1,\foldAngle_2,\foldAngle_3,\foldAngle_2,\foldAngle_1,\foldAngle_2,\foldAngle_3,\foldAngle_2\right)$.

As a simple example of this generalization, consider searching for compatible, folded states of the symmetric $8$-fold vertex such that $\genAngle = \pi / 2$--which is the most symmetry for folded configurations that the vertex admits: $\dihedral{4}$.
In this case, $k = 2$ and $\Facet{c} = \Facet{1}$.
Using \eqref{eq:general-reflection-consistency-condition} and \eqref{eq:general-reflection-rotation-other-fold-angle}, one obtains
\begin{equation} \label{eq:waterbomb-fold-angles}
	\left(\foldAngle_1, \foldAngle_2\right) = \begin{cases}
		\left(2 \arccos \left(\frac{
			\left|\sqrt{2} + 2 \left(1 - 2\heaviside{\elevAngle}\right) \cos \elevAngle\right|
		}{
			2 + \sqrt{2}\left(1 - 2\heaviside{\elevAngle}\right) \cos \elevAngle
	}\right), -2\left|\elevAngle\right|\right), & \foldAngle_1 \text{ valley, } \foldAngle_2 \text{ mountain}\\
	\left(-2 \arccos \left(\frac{
		\left|\sqrt{2} - 2 \left(1 - 2\heaviside{\elevAngle}\right) \cos \elevAngle\right|
	}{
		2 - \sqrt{2}\left(1 - 2\heaviside{\elevAngle}\right) \cos \elevAngle
	}\right), 2\left|\elevAngle\right|\right), & \text{vice versa}
	\end{cases},
\end{equation}
where $\heaviside{x} = \left(x + \left|x\right|\right) / 2 x$, is the Heaviside step function.

The last step then is to make the assignments $\left(\foldAngle_1, \dots, \foldAngle_8\right) = \left(\foldAngle_1, \foldAngle_2, \foldAngle_1, \foldAngle_2, \foldAngle_1, \foldAngle_2, \foldAngle_1, \foldAngle_2\right)$.
This recovers the well-known solution for the ``fully-symmetric'' folding of the $8$-fold waterbomb where the crease assignments are alternating mountain and valley~\cite{grasinger2022multistability}.
For the first solution in \eqref{eq:waterbomb-fold-angles}, the odd creases are valley and the even creases are mountain; for the second, vice versa.
More details on this kinematic description of the $8$-fold waterbomb, its multistability when the creases store elastic energy, and various other interesting properties can be found in~\cite{grasinger2022multistability,hanna2014waterbomb,brunck2016elastic}.

Notice that, in assigning fold angles around the vertex, the sequence of indices on the right side of the assignment were increasing twice for the $\dihedral{2}$ folding and four times for the $\dihedral{4}$ folding;
in addition, for each of the foldings with $\dihedral{1}$ folding (i.e. a single reflection symmetry), the sequence of indices is increasing once.
In general, the sequence of indices for $\dihedral{2 h}$ foldings will increase $2 h$ times.
Similarly, $2 h$ is the number of times that the condition \eqref{eq:general-reflection-rotation-consistency-condition} is needed to be substituted into \eqref{eq:compatibility} in order to show that the reduced order solution does indeed satisfy compatibility.
This, in addition to the introduction of the parameter $\elevAngle$, all corresponds with a reduction of degrees of freedom from $\NCreases - 3$ to $\NCreases / m$, where $m$ is the smallest integer such that $\rotAbout{\ethree}^m\left(\genAngle\right) = \Iden$, i.e. $m$ is the order of the generating rotation.

\subsection{General $8$-vertex with $\dihedral{2}$ symmetry} \label{sec:8G-rotation-reflection}

\begin{figure}
	\centering
	\includegraphics[width=\linewidth]{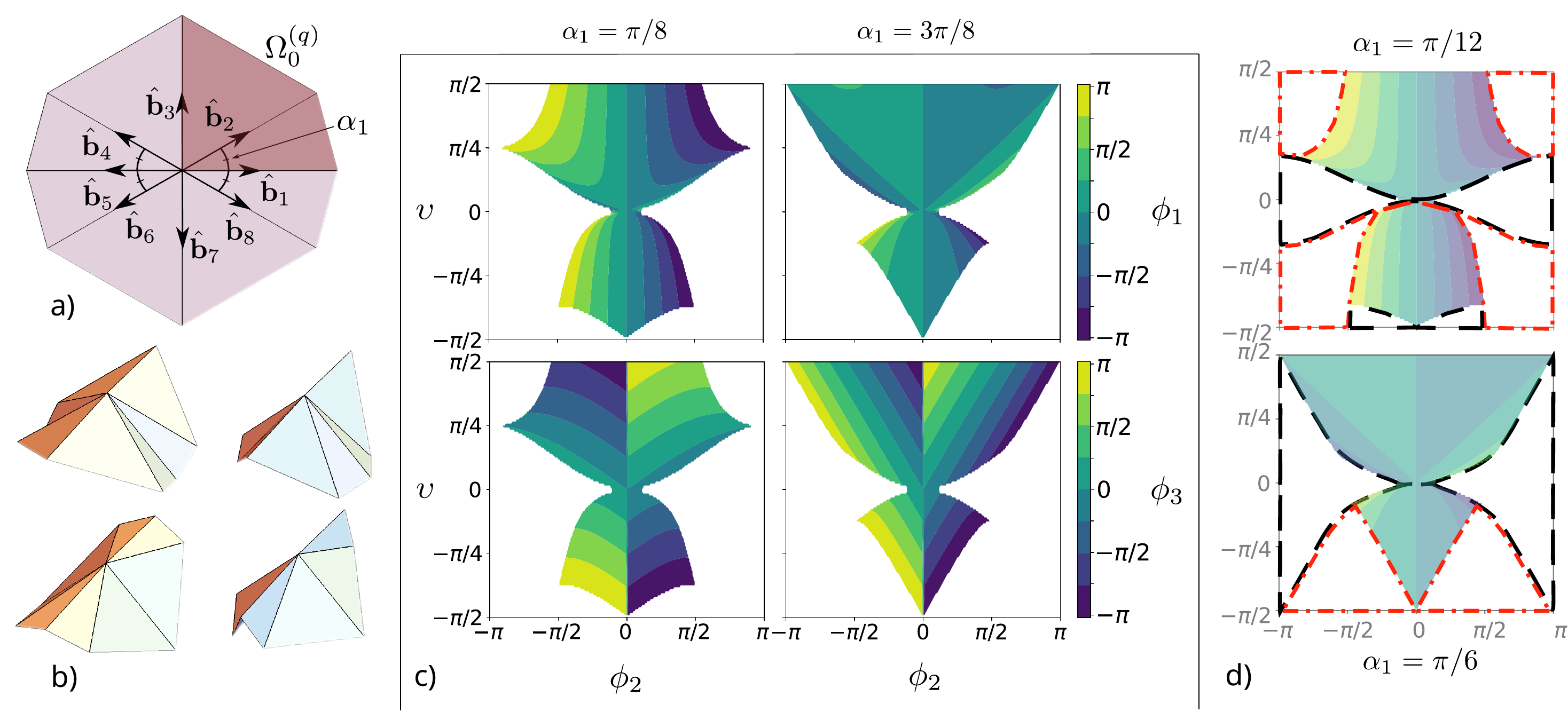}
	\caption{
		\textbf{Kinematics of a degree $8$ vertex with $\dihedral{2}$ symmetry and variable $\sectAngle_1$.}
		a) Geometry of degree $8$ vertex with $\dihedral{2}$ symmetry parameterized by $\sectAngle_1$.
		b) Example folded configurations for $\sectAngle_1 = \pi / 8$ (top row) and $\sectAngle_1 = 3 \pi / 8$ (bottom row).
		c) Contours of $\foldAngle_1 = \foldAngle_1\left(\elevAngle, \foldAngle_2\right)$ (top row) and $\foldAngle_3 = \foldAngle_3\left(\elevAngle, \foldAngle_2\right)$ (bottom row) for $\sectAngle_1 = \pi / 8$ (left column) and $\sectAngle_1 = 3\pi/8$ (right column).
		d) The admissible part of the $\left(\elevAngle, \foldAngle_2\right)$ for $\sectAngle = \pi / 12$ (top) and $\sectAngle_1 = \pi / 6$ (bottom) where inadmissible regions are delineated by a black dashed line when no solution exists and a red dash-dot line when the region is inadmissible due to contact. 
        More of the domain has a folding solution as $\left|\sectAngle_1 - \pi / 4\right|$ increases, but more of the domain is also inadmissible due to contact.
	}
	\label{fig:gen-8-setup-and-results}
\end{figure}

\newcommand{\auxJ}{\mathcal{J}}
\newcommand{\auxK}{\mathcal{K}}
As a simple example of this generalization, an $8$-fold vertex such that $\sectAngle_1 = \sectAngle, \sectAngle_2 = \pi / 2 - \sectAngle$, and $\group = \genBy{\refl{\etwo}, \rotate{\ethree}{\pi / 2}}$, as shown in \fref{fig:gen-8-setup-and-results}.a. 
Then, using \eqref{eq:general-reflection-rotation-consistency-condition}, we obtain
\begin{equation}
\partFoldAngle_1 = \left\{
   \begin{array}{r}
   \left\{
   \begin{array}{rrr}
    \arctan\Bigg(& \auxK \cos\sectAngle \left(\cos\foldAngle_2-1\right) - \auxJ\left(\cos^2\sectAngle \cos\foldAngle_2 + \sin^2\sectAngle\right), \: & \\
    & \cos\sectAngle \sin\foldAngle_2 \auxJ - \auxK \left(\cos^2\sectAngle \cos\foldAngle_2 + \sin^2\sectAngle\right) \tan\left(\frac{\foldAngle_2}{2}\right) & \Bigg), \quad \foldAngle_2 > 0 \\
     -\arctan\Bigg(& -\auxK \cos\sectAngle \left(\cos\foldAngle_2-1\right) - \auxJ\left(\cos^2\sectAngle \cos\foldAngle_2 + \sin^2\sectAngle\right), \: & \\
    & -\cos\sectAngle \sin\foldAngle_2 \auxJ - \auxK \left(\cos^2\sectAngle \cos\foldAngle_2 + \sin^2\sectAngle\right) \tan\left(\frac{\foldAngle_2}{2}\right) & \Bigg), \quad \foldAngle_2 < 0
   \end{array} 
   \right. 
   \\
   \left\{
   \begin{array}{rrr}
      -\arctan\Bigg(& \auxK \cos\sectAngle \left(\cos\foldAngle_2-1\right) + \auxJ\left(\cos^2\sectAngle \cos\foldAngle_2 + \sin^2\sectAngle\right), \: & \\ 
       & \cos\sectAngle \sin\foldAngle_2 \auxJ + \auxK \left(\cos^2\sectAngle \cos\foldAngle_2 + \sin^2\sectAngle\right) \tan\left(\frac{\foldAngle_2}{2}\right) & \Bigg), \quad \foldAngle_2 > 0 \\
     \arctan\Bigg(& -\auxK \cos\sectAngle \left(\cos\foldAngle_2-1\right) + \auxJ\left(\cos^2\sectAngle \cos\foldAngle_2 + \sin^2\sectAngle\right), \: & \\
      & -\cos\sectAngle \sin\foldAngle_2 \auxJ + \auxK \left(\cos^2\sectAngle \cos\foldAngle_2 + \sin^2\sectAngle\right) \tan\left(\frac{\foldAngle_2}{2}\right) & \Bigg), \quad \foldAngle_2 < 0
      \end{array} 
      \right.
   \end{array} \right.
\end{equation}
where
\begin{equation}
\begin{split}
\auxJ\left(\elevAngle, \foldAngle_2\right) &\coloneqq \sqrt{5 - 8\cos\elevAngle + 4\cos\foldAngle_2 - \cos\left(2\foldAngle_2\right) + 8\cos\left(4\sectAngle\right) \sin^4\left(\frac{\foldAngle_2}{2}\right)}, \\
\auxK\left(\elevAngle, \foldAngle_2\right) &\coloneqq 4\left|\cos\sectAngle \sin\elevAngle \sin\foldAngle_2\right|\sin\sectAngle \cot\elevAngle,
\end{split}
\end{equation}
and, consequently, the solution is only real valued (and therefore valid) provided: $\auxJ \in \Reals$.
Example folded states for $\elevAngle = \pi / 4, \foldAngle_2 = \pi / 3$ and $\sectAngle_1 = \pi / 8$ (top row) and $\sectAngle_1 = 3 \pi / 8$ (bottom row) for the first (left column) and second (right column) folding branches are shown in \fref{fig:gen-8-setup-and-results}.b.
Let $\foldAngle_1 = \convToFoldAngle\left(2 \partFoldAngle_1\left(\elevAngle, \foldAngle_2\right)\right)$.
The contours for $\foldAngle_1$ (top row) and $\foldAngle_3 = \foldAngle_3\left(\elevAngle, \foldAngle_2\right)$ (bottom row) are shown in \fref{fig:gen-8-setup-and-results}.c for $\sectAngle_1 = \pi / 8$ (left column) and $\sectAngle_1 = 3 \pi / 8$ (right column).
There are several notable features about the $\dihedral{2}$ kinematics of the degree 8 vertex with varying $\sectAngle_1$:
\begin{enumerate}[1)]
  \item When $\sectAngle_1 < \pi / 4$, creases $\set{1, 2, 8}$ and $\set{3, 4, 5}$ are nearer to each other than either of collinear creases $3$ and $7$.
    The left column of \fref{fig:gen-8-setup-and-results}.c ($\sectAngle_1 = \pi / 8$) shows that, when creases $1$ and $2$ are nearer to each other, $\foldAngle_1$ folds primarily as a function of $\foldAngle_2$ and is almost independent of the elevation angle, $\elevAngle$.
    In contrast, $\foldAngle_3$ has a stronger dependence on $\elevAngle$ and only folds flat as $|\elevAngle| \rightarrow \pi / 2$.
  \item Alternatively, when $\sectAngle_1 > \pi / 4$, creases $\set{2, 3, 4}$ and $\set{6, 7, 8}$ are nearer to each other than either of collinear creases $1$ and $5$.
    By the right column (and top row) of \fref{fig:gen-8-setup-and-results}.c ($\sectAngle_1 = 3 \pi / 8$), one can see that $\foldAngle_1$ folds less as a function of $\elevAngle$ and $\foldAngle_2$; i.e. $\left|\foldAngle_1\right| \leq \pi / 2$ for much of the $\left(\elevAngle, \foldAngle_2\right) \in \left[-\pi / 2, \pi / 2\right] \times \left[-\pi, \pi\right]$ domain.
    In contrast, the contours of $\foldAngle_3$ take the form of thin, linear bands with slope of unity (i.e. along lines of $\elevAngle = \foldAngle_2 + \text{const.}$)--which suggests that $\foldAngle_3$ folds more as a function of $\left(\elevAngle, \foldAngle_2\right)$ when $\sectAngle_1 \rightarrow \pi / 2$.
    See \fref{fig:gen-8-setup-and-results}, for example.
  \item \Fref{fig:gen-8-setup-and-results}.d shows inadmissible regions are delineated by a black dashed line when no solution exists and a red dash-dot line when the region is inadmissible due to contact. 
  Note that more of the domain has a folding solution as $\left|\sectAngle_1 - \pi / 4\right|$ increases.
    In other words, the condition $\auxJ \in \Reals$ (i.e. $5 - 8\cos\elevAngle + 4\cos\foldAngle_2 - \cos\left(2\foldAngle_2\right) + 8\cos\left(4\sectAngle\right) \sin^4\left(\frac{\foldAngle_2}{2}\right) \geq 0$) is met for a greater part of the domain as $\left|\sectAngle_1 - \pi / 4\right|$ increases.
    Physically, as creases in the vertex either collect nearer to $1$ and $4$ ($\sectAngle_1 < \pi / 4$) or nearer to $3$ and $7$ ($\sectAngle_1 > \pi / 4$), the center vertex does not need to flex up or down as much in order to allow folding about crease $2$.
    In terms of contact, more of the domain becomes inadmissible as $\left|\sectAngle_1 - \pi / 4\right|$.
    This may be due, in part, to the fact that the minimum distances between pairs of faces in the flat state is decreasing (i.e. there is a smaller minimum sector angle).
    An animation of an $\sectAngle_1 = \pi / 6$ vertex folding through 1) configuration space into admissible regions, 2) regions where no solution exists, and 3) regions where contact occurs, is shown in Animation \#5 included in the ESI.
\end{enumerate}

\section{Conclusion} \label{sec:conclusion}
The Lagrangian approach to origami makes the deformation map the fundamental kinematic description of the origami.
Here we showed its advantages toward exploiting symmetry via reduced order compatibility conditions.
The reduced order conditions were solved exactly for multi-degree of freedom kinematics of symmetric degree $6$ and degree $8$ vertices.  
It is expected that the exact solutions will facilitate and inspire new origami design algorithms. While the results herein were restricted to single vertices with an even number of creases, the developed approach should generalize readily to multivertex crease patterns and odd degree vertices.
For instance, globally compatible solutions for a multivertex crease pattern can be obtained by solving for local compatibility at each of the vertices and then enforcing that, where vertices share a crease, the corresponding fold angle agrees.
For odd degree vertices, one could, in principle, add a ``pseudo'' crease that (agrees with the symmetry of interest and) results in an even degree vertex that has a reduced order compatibility condition already outlined in this work.
Then the odd degree vertex solution could be obtained by subsequently solving for the fold angle of the pseudo crease as zero.
While these generalizations are straight forward, it is possible that different generalizations can be made which are more amenable to exact solution or more computationally efficient, which presents an interesting topic for future work.

In this study, we limited our analysis to origami structures with negligible thickness.  Finite thickness origami structures are an important class of foldable systems with many practical advantages.  Novel hinge concepts \cite{yellowhorse2018creating, lang2018review} and fold topology strategies \cite{chen2015origami, zhu2024large} are effective approaches to address this challenge. One of the consequences of thick systems is that self-contact at the hinge constrains the kinematic range of folding, often only allowing folding in one direction from the flat state. In light of this, incorporating finite thickness into our analysis would likely reduce the domain of accessible folding kinematics we presented. Extending our Lagrangian formulation to accommodate thickness is another potential direction for future study.

\section*{Acknowledgements}
We are grateful to Richard James and Kaushik Bhattacharya for many insightful discussions.
We acknowledge the support of the Laboratory University Collaboration Initiative sponsored by the US Office of the Secretary of Defence and the support of the Air Force Research Laboratory.

\appendix
\section{Algebraic perspective of the $4$-vertex with reflection symmetry}
\label{app:algebraic-4-vertex}

For the $4$ vertex with reflection symmetry, \eqref{eq:group-compatibility} takes the form
\begin{equation}
	\rotate{\CreaseVec_1}{\foldAngle_1} \rotate{\CreaseVec_2}{\foldAngle_2} \rotate{\CreaseVec_3}{\foldAngle_3} \rotate{\refl{\etwo} \CreaseVec_2}{\foldAngle_4} = \Iden.
\end{equation}
Using \eqref{eq:rotations-inverse}, \eqref{eq:rotations-about-transformed-axis}, and \eqref{eq:rotation-reflection-about-ortho-axes}, this can be rewritten as
\begin{equation} \label{eq:intermediate-4-vertex-1}
\rotate{\CreaseVec_1}{\frac{\foldAngle_1}{2}} \rotate{\CreaseVec_2}{\foldAngle_2} \rotate{\CreaseVec_3}{\foldAngle_3} \refl{\etwo} \left( \rotate{\CreaseVec_2}{-\foldAngle_4} \rotate{\CreaseVec_1}{-\frac{\foldAngle_1}{2}} \right)\refl{\etwo} = \Iden.
\end{equation}
Notice that, if one can work the first part of the product, namely $\rotate{\CreaseVec_1}{\foldAngle_1 / 2} \rotate{\CreaseVec_2}{\foldAngle_2}$, in between the $\refl{\etwo}$ and $\refl{\etwo}$ transformations, then terms cancel through appropriate choice of $\foldAngle_4$.
To accomplish this, recall the notion of constructing a compatible folded state as the orbit of $\facet{h}$ (\fref{fig:4-vertex}).
In this case, by construction, $\facet{3} = \refl{\etwo} \facet{2}$ and $\facet{4} = \refl{\etwo} \facet{1}$.
This implies that
\begin{subequations}
	\label{eq:consistency-4-vertex}
	\begin{align}
	    \label{eq:consistency-4-vertex-23}
		\rotate{\CreaseVec_1}{\frac{\foldAngle_1}{2}} \rotate{\CreaseVec_2}{\foldAngle_2} \rotate{\CreaseVec_3}{\foldAngle_3} &= \refl{\etwo} \rotate{\CreaseVec_1}{\frac{\foldAngle_1}{2}} \rotate{\CreaseVec_2}{\foldAngle_2} \refl{\etwo}, \\
		\label{eq:consistency-4-vertex-14}
		\rotate{\CreaseVec_1}{\frac{\foldAngle_1}{2}} \rotate{\CreaseVec_2}{\foldAngle_2} \rotate{\CreaseVec_3}{\foldAngle_3} \rotate{\CreaseVec_4}{\foldAngle_4} &= \rotate{\CreaseVec_1}{-\frac{\foldAngle_1}{2}} = \refl{\etwo} \rotate{\CreaseVec_1}{\frac{\foldAngle_1}{2}} \refl{\etwo},
	\end{align}
\end{subequations}
respectively.
The second condition, \eqref{eq:consistency-4-vertex-14}, is trivially satisfied (by \eqref{eq:rotation-reflection-about-ortho-axes}).
The first condition, \eqref{eq:consistency-4-vertex-23}, can be satisfied by proper choice of $\foldAngle_1$, $\foldAngle_2$ and $\foldAngle_3$.
Assuming \eqref{eq:consistency-4-vertex-23} holds, we see that \eqref{eq:intermediate-4-vertex-1} takes the form
\begin{equation}
	\rotate{\CreaseVec_1}{\frac{\foldAngle_1}{2}} \rotate{\CreaseVec_2}{\foldAngle_2 - \foldAngle_4} \rotate{\CreaseVec_1}{-\frac{\foldAngle_1}{2}} = \Iden,
\end{equation}
which is satisfied if and only if $\foldAngle_4 = \foldAngle_2$, consistent with the assumed symmetry of the folded configuration.
Thus, \eqref{eq:consistency-4-vertex-23} represents a dimensionally reduced form of \eqref{eq:compatibility}.
By \eqref{eq:rotation-reflection-about-ortho-axes}, the ansatz $\rotate{\CreaseVec_1}{\foldAngle_1 / 2} \rotate{\CreaseVec_2}{\foldAngle_2} = \rotate{\etwo}{\genAngle} \rotate{\CreaseVec_3}{\genAngle'}$ can be used to readily solve \eqref{eq:consistency-4-vertex-23}--which is equivalent to the condition given in \eqref{eq:reduced-4-vertex} that was developed from the geometric perspective.
Upon using this ansatz and then backward substitution, one finds that $\foldAngle_3 = -\foldAngle_1$.


\section{Self-contact detection procedure}
\label{app:self-contact}
Solutions to the compatibility conditions formulated in this work (and others) do not consider self-contact.
Detecting contact can be expensive~\cite{zhu2019efficient}, but we find here that the Lagrangian approach--along with analytical solutions--provides a path forward.
Using the deformation map, one can readily determine the locations of the facets in the folded configuration.
Intersection of a line with a triangle, in $3$ dimensions, can be checked for in a straight forward manner.
Let $\lpt_1$ and $\lpt_2$ denote the end points of the line and $\tpt_1$, $\tpt_2$, and $\tpt_3$ denote the vertices of the triangle.
Given $4$ points, $\xC_1, \dots, \xC_4$, the signed volume of the tetrahedron determined by the points is
\begin{equation}
	\sVol\left(\xC_1, \xC_2, \xC_3, \xC_4\right) = \frac{1}{6} \left( \left(\xC_2 - \xC_1\right) \times \left(\xC_3 - \xC_1\right) \right) \cdot \left(\xC_4 - \xC_1\right).
\end{equation}
If the $\sVol\left(\lpt_1, \tpt_1, \tpt_2, \tpt_3\right)$ and $\sVol\left(\lpt_2, \tpt_1, \tpt_2, \tpt_3\right)$ are \emph{different} signs, then the line passes through the plane of the triangle; and if
	$\sVol\left(\lpt_1, \lpt_2, \tpt_1, \tpt_2\right), \: \sVol\left(\lpt_1, \lpt_2, \tpt_2, \tpt_3\right), \: \sVol\left(\lpt_1, \lpt_2, \tpt_3, \tpt_1\right),$
are all the same sign, then the line lies in the (infinite) extrusion of the triangle normal to its plane.
Thus, if both conditions are met, the line and triangle intersect.
For instance, to check for self-contact in the $6$ vertex with reflection symmetry, we consider the following
\begin{enumerate}[1)]
	\item if any $\creaseVec_{i} \cdot \etwo < 0$ for $i = 1, \dots, 4$, then, by symmetry, self-contact has occurred;
	\item each pair of lines and facets in $\facet{1} \cup \facet{2} \cup \facet{3}$ are checked for intersection by signed volumes of the corresponding tetrahedrons.
	To be precise, there is an intersection between a facet and facet edge pair if and only if
	\begin{equation}
	\begin{split}
		\sgn\left(\sVol\left(\lpt_1, \tpt_1, \tpt_2, \tpt_3\right)\right) &\neq \sgn\left(\sVol\left(\lpt_2, \tpt_1, \tpt_2, \tpt_3\right)\right) \quad \text { and } \\
		\sgn\left(\sVol\left(\lpt_1, \lpt_2, \tpt_1, \tpt_2\right)\right) &= \sgn\left(\sVol\left(\lpt_1, \lpt_2, \tpt_2, \tpt_3\right)\right) = \sgn\left(\sVol\left(\lpt_1, \lpt_2, \tpt_3, \tpt_1\right)\right).
	\end{split}
	\end{equation}
	If an intersection between any pair of lines and facets occur, we say contact has occurred and the configuration is inadmissible.
\end{enumerate}
Alternative algorithms for detecting contact that are likely more efficient can be found in~\cite{moller2005fast,zhu2019efficient}.
Despite the relative expense of the contact problem, we can still efficiently map out the admissible regions of configuration space  where the structure is folded symmetrically in minutes (as opposed to hours or days). 
This efficiency comes from several factors
\begin{inparaenum}[1)]
\item we have closed form solutions for the fold angles in hand,
\item we can directly obtain the deformation map from the Lagrangian approach and fold angles, and
\item the symmetry of the deformed configurations simplifies the contact problem.
\end{inparaenum}

When self-contact occurs will depend on the geometry of the facets of the origami.
In this work, all facet geometries are chosen such that all of the creases are the same length, and outer edges are added between neighboring creases to form triangular facets.
Although the procedure outlined herein is specific to triangular facets, it can readily be extended to facets of general shape by first approximating the shape of each facet using a triangulation procedure.

\bibliographystyle{unsrtnat}
\bibliography{master}

\begin{thebibliography}{81}
\providecommand{\natexlab}[1]{#1}
\providecommand{\url}[1]{\texttt{#1}}
\expandafter\ifx\csname urlstyle\endcsname\relax
  \providecommand{\doi}[1]{doi: #1}\else
  \providecommand{\doi}{doi: \begingroup \urlstyle{rm}\Url}\fi

\bibitem[Schenk et~al.(2014)Schenk, Viquerat, Seffen, and
  Guest]{schenk2014review}
Mark Schenk, Andrew~D Viquerat, Keith~A Seffen, and Simon~D Guest.
\newblock Review of inflatable booms for deployable space structures: packing
  and rigidization.
\newblock \emph{Journal of Spacecraft and Rockets}, 51\penalty0 (3):\penalty0
  762--778, 2014.

\bibitem[Zirbel et~al.(2013)Zirbel, Lang, Thomson, Sigel, Walkemeyer, Trease,
  Magleby, and Howell]{zirbel2013accommodating}
Shannon~A Zirbel, Robert~J Lang, Mark~W Thomson, Deborah~A Sigel, Phillip~E
  Walkemeyer, Brian~P Trease, Spencer~P Magleby, and Larry~L Howell.
\newblock Accommodating thickness in origami-based deployable arrays.
\newblock \emph{Journal of Mechanical Design}, 135\penalty0 (11), 2013.

\bibitem[Verzoni and Rais-Rohani(2022)]{verzoni2022transition}
Anthony Verzoni and Masoud Rais-Rohani.
\newblock Transition analysis of flat-foldable origami-inspired deployable
  shelter concepts.
\newblock \emph{Engineering Structures}, 273:\penalty0 115074, 2022.

\bibitem[Melancon et~al.(2021)Melancon, Gorissen, Garc{\'\i}a-Mora, Hoberman,
  and Bertoldi]{melancon2021multistable}
David Melancon, Benjamin Gorissen, Carlos~J Garc{\'\i}a-Mora, Chuck Hoberman,
  and Katia Bertoldi.
\newblock Multistable inflatable origami structures at the metre scale.
\newblock \emph{Nature}, 592\penalty0 (7855):\penalty0 545--550, 2021.

\bibitem[Kuribayashi et~al.(2006)Kuribayashi, Tsuchiya, You, Tomus, Umemoto,
  Ito, and Sasaki]{kuribayashi2006self}
Kaori Kuribayashi, Koichi Tsuchiya, Zhong You, Dacian Tomus, Minoru Umemoto,
  Takahiro Ito, and Masahiro Sasaki.
\newblock Self-deployable origami stent grafts as a biomedical application of
  ni-rich tini shape memory alloy foil.
\newblock \emph{Materials Science and Engineering: A}, 419\penalty0
  (1-2):\penalty0 131--137, 2006.

\bibitem[Andersen et~al.(2009)Andersen, Dong, Nielsen, Jahn, Subramani,
  Mamdouh, Golas, Sander, Stark, Oliveira, Pedersen, Birkedal, Besenbacher,
  Gothelf, and Kjems]{andersen2009self}
Ebbe~S Andersen, Mingdong Dong, Morten~M Nielsen, Kasper Jahn, Ramesh
  Subramani, Wael Mamdouh, Monika~M Golas, Bjoern Sander, Holger Stark,
  Cristiano~LP Oliveira, Jan~S Pedersen, Victoria Birkedal, Flemming
  Besenbacher, Kurt Gothelf, and J{\o}rgen Kjems.
\newblock Self-assembly of a nanoscale {DNA} box with a controllable lid.
\newblock \emph{Nature}, 459\penalty0 (7243):\penalty0 73--76, 2009.

\bibitem[Velvaluri et~al.(2021)Velvaluri, Soor, Plucinsky, de~Miranda, James,
  and Quandt]{velvaluri2021origami}
Prasanth Velvaluri, Arun Soor, Paul Plucinsky, Rodrigo~Lima de~Miranda,
  Richard~D James, and Eckhard Quandt.
\newblock Origami-inspired thin-film shape memory alloy devices.
\newblock \emph{Scientific reports}, 11\penalty0 (1):\penalty0 10988, 2021.

\bibitem[Jiang et~al.(2019)Jiang, Liu, Liu, Wang, and
  Ding]{jiang2019rationally}
Qiao Jiang, Shaoli Liu, Jianbing Liu, Zhen-Gang Wang, and Baoquan Ding.
\newblock Rationally designed dna-origami nanomaterials for drug delivery in
  vivo.
\newblock \emph{Advanced Materials}, 31\penalty0 (45):\penalty0 1804785, 2019.

\bibitem[Liu et~al.(2023)Liu, Loh, Siti, Too, Anderson, and Wang]{liu2023light}
Xiao~Rui Liu, Iong~Ying Loh, Winna Siti, Hon~Lin Too, Tommy Anderson, and
  Zhisong Wang.
\newblock A light-operated integrated dna walker--origami system beyond bridge
  burning.
\newblock \emph{Nanoscale Horizons}, 8\penalty0 (6):\penalty0 827--841, 2023.

\bibitem[Cho et~al.(2011)Cho, Keung, Verellen, Lagae, Moshchalkov, Van~Dorpe,
  and Gracias]{cho2011nanoscale}
Jeong-Hyun Cho, Michael~D Keung, Niels Verellen, Liesbet Lagae, Victor
  Moshchalkov, Pol Van~Dorpe, and David~H Gracias.
\newblock Nanoscale origami for 3d optics.
\newblock \emph{Small}, 7\penalty0 (14):\penalty0 1943--1948, 2011.

\bibitem[Grasinger and Sharma(2024)]{grasingerIPnanoscale}
Matthew Grasinger and Pradeep Sharma.
\newblock Thermal fluctuations (eventually) unfold nanoscale origami.
\newblock \emph{Journal of the Mechanics and Physics of Solids}, page 105527,
  2024.
\newblock ISSN 0022-5096.

\bibitem[Novelino et~al.(2020)Novelino, Ze, Wu, Paulino, and
  Zhao]{novelino2020untethered}
Larissa~S Novelino, Qiji Ze, Shuai Wu, Glaucio~H Paulino, and Ruike Zhao.
\newblock Untethered control of functional origami microrobots with distributed
  actuation.
\newblock \emph{Proceedings of the National Academy of Sciences}, 117\penalty0
  (39):\penalty0 24096--24101, 2020.

\bibitem[Wu et~al.(2021)Wu, Ze, Dai, Udipi, Paulino, and
  Zhao]{wu2021stretchable}
Shuai Wu, Qiji Ze, Jize Dai, Nupur Udipi, Glaucio~H Paulino, and Ruike Zhao.
\newblock Stretchable origami robotic arm with omnidirectional bending and
  twisting.
\newblock \emph{Proceedings of the National Academy of Sciences}, 118\penalty0
  (36), 2021.

\bibitem[Sadeghi et~al.(2021)Sadeghi, Allison, Bestill, and Li]{sadeghi2021tmp}
Sahand Sadeghi, Samuel~R Allison, Blake Bestill, and Suyi Li.
\newblock Tmp origami jumping mechanism with nonlinear stiffness.
\newblock \emph{Smart Materials and Structures}, 30\penalty0 (6):\penalty0
  065002, 2021.

\bibitem[Bhovad et~al.(2019)Bhovad, Kaufmann, and Li]{bhovad2019peristaltic}
Priyanka Bhovad, Joshua Kaufmann, and Suyi Li.
\newblock Peristaltic locomotion without digital controllers: Exploiting
  multi-stability in origami to coordinate robotic motion.
\newblock \emph{Extreme Mechanics Letters}, 32:\penalty0 100552, 2019.

\bibitem[Li et~al.(2016)Li, Fang, and Wang]{li2016recoverable}
S~Li, H~Fang, and KW~Wang.
\newblock Recoverable and programmable collapse from folding pressurized
  origami cellular solids.
\newblock \emph{Physical Review Letters}, 117\penalty0 (11):\penalty0 114301,
  2016.

\bibitem[Addis et~al.(2023)Addis, Rojas, and Arrieta]{addis2023connecting}
Clark~C Addis, Salvador Rojas, and Andres~F Arrieta.
\newblock Connecting the branches of multistable non-euclidean origami by
  crease stretching.
\newblock \emph{Physical Review E}, 108\penalty0 (5):\penalty0 055001, 2023.

\bibitem[Treml et~al.(2018)Treml, Gillman, Buskohl, and Vaia]{treml2018origami}
Benjamin Treml, Andrew Gillman, Philip Buskohl, and Richard Vaia.
\newblock Origami mechanologic.
\newblock \emph{Proceedings of the National Academy of Sciences}, 115\penalty0
  (27):\penalty0 6916--6921, 2018.

\bibitem[Jules et~al.(2022)Jules, Reid, Daniels, Mungan, and
  Lechenault]{jules2022delicate}
Th{\'e}o Jules, Austin Reid, Karen~E Daniels, Muhittin Mungan, and
  Fr{\'e}d{\'e}ric Lechenault.
\newblock Delicate memory structure of origami switches.
\newblock \emph{Physical Review Research}, 4\penalty0 (1):\penalty0 013128,
  2022.

\bibitem[Bhovad and Li(2021)]{bhovad2021physical}
Priyanka Bhovad and Suyi Li.
\newblock Physical reservoir computing with origami and its application to
  robotic crawling.
\newblock \emph{Scientific Reports}, 11\penalty0 (1):\penalty0 1--18, 2021.

\bibitem[Boatti et~al.(2017)Boatti, Vasios, and Bertoldi]{boatti2017origami}
Elisa Boatti, Nikolaos Vasios, and Katia Bertoldi.
\newblock Origami metamaterials for tunable thermal expansion.
\newblock \emph{Advanced Materials}, 29\penalty0 (26):\penalty0 1700360, 2017.

\bibitem[Liu et~al.(2022)Liu, Pratapa, Misseroni, Tachi, and
  Paulino]{liu2022triclinic}
Ke~Liu, Phanisri~P Pratapa, Diego Misseroni, Tomohiro Tachi, and Glaucio~H
  Paulino.
\newblock Triclinic metamaterials by tristable origami with reprogrammable
  frustration.
\newblock \emph{Advanced Materials}, 34\penalty0 (43):\penalty0 2107998, 2022.

\bibitem[Miyazawa et~al.(2021)Miyazawa, Yasuda, Kim, Lynch, Tsujikawa,
  Kunimine, Raney, and Yang]{miyazawa2021heterogeneous}
Yasuhiro Miyazawa, Hiromi Yasuda, Hyungkyu Kim, James~H Lynch, Kosei Tsujikawa,
  Takahiro Kunimine, Jordan~R Raney, and Jinkyu Yang.
\newblock Heterogeneous origami-architected materials with variable stiffness.
\newblock \emph{Communications Materials}, 2\penalty0 (1):\penalty0 1--7, 2021.

\bibitem[Misseroni et~al.(2022)Misseroni, Pratapa, Liu, and
  Paulino]{misseroni2022experimental}
Diego Misseroni, Phanisri~P Pratapa, Ke~Liu, and Glaucio~H Paulino.
\newblock Experimental realization of tunable poisson’s ratio in deployable
  origami metamaterials.
\newblock \emph{Extreme Mechanics Letters}, 53:\penalty0 101685, 2022.

\bibitem[Zhai et~al.(2020)Zhai, Wang, Lin, Wu, and Jiang]{zhai2020situ}
Zirui Zhai, Yong Wang, Ken Lin, Lingling Wu, and Hanqing Jiang.
\newblock In situ stiffness manipulation using elegant curved origami.
\newblock \emph{Science advances}, 6\penalty0 (47):\penalty0 eabe2000, 2020.

\bibitem[Silverberg et~al.(2014)Silverberg, Evans, McLeod, Hayward, Hull,
  Santangelo, and Cohen]{silverberg2014using}
Jesse~L Silverberg, Arthur~A Evans, Lauren McLeod, Ryan~C Hayward, Thomas Hull,
  Christian~D Santangelo, and Itai Cohen.
\newblock Using origami design principles to fold reprogrammable mechanical
  metamaterials.
\newblock \emph{science}, 345\penalty0 (6197):\penalty0 647--650, 2014.

\bibitem[Schenk and Guest(2013)]{schenk2013geometry}
Mark Schenk and Simon~D Guest.
\newblock Geometry of miura-folded metamaterials.
\newblock \emph{Proceedings of the National Academy of Sciences}, 110\penalty0
  (9):\penalty0 3276--3281, 2013.

\bibitem[Liu et~al.(2018)Liu, Silverberg, Evans, Santangelo, Lang, Hull, and
  Cohen]{liu2018topological}
Bin Liu, Jesse~L Silverberg, Arthur~A Evans, Christian~D Santangelo, Robert~J
  Lang, Thomas~C Hull, and Itai Cohen.
\newblock Topological kinematics of origami metamaterials.
\newblock \emph{Nature Physics}, 14\penalty0 (8):\penalty0 811--815, 2018.

\bibitem[Grasinger et~al.(2022)Grasinger, Gillman, and
  Buskohl]{grasinger2022multistability}
Matthew Grasinger, Andrew Gillman, and Philip~R Buskohl.
\newblock Multistability, symmetry and geometric conservation in eightfold
  waterbomb origami.
\newblock \emph{Proceedings of the Royal Society A}, 478\penalty0
  (2268):\penalty0 20220270, 2022.

\bibitem[Pratapa et~al.(2019)Pratapa, Liu, and Paulino]{pratapa2019geometric}
Phanisri~P Pratapa, Ke~Liu, and Glaucio~H Paulino.
\newblock Geometric mechanics of origami patterns exhibiting poisson’s ratio
  switch by breaking mountain and valley assignment.
\newblock \emph{Physical Review Letters}, 122\penalty0 (15):\penalty0 155501,
  2019.

\bibitem[Brunck et~al.(2016)Brunck, Lechenault, Reid, and
  Adda-Bedia]{brunck2016elastic}
V~Brunck, F~Lechenault, A~Reid, and M~Adda-Bedia.
\newblock Elastic theory of origami-based metamaterials.
\newblock \emph{Physical Review E}, 93\penalty0 (3):\penalty0 033005, 2016.

\bibitem[Sessions et~al.(2019)Sessions, Cook, Fuchi, Gillman, Huff, and
  Buskohl]{sessions2019origami}
Deanna Sessions, Alexander Cook, Kazuko Fuchi, Andrew Gillman, Gregory Huff,
  and Philip Buskohl.
\newblock Origami-inspired frequency selective surface with fixed frequency
  response under folding.
\newblock \emph{Sensors}, 19\penalty0 (21):\penalty0 4808, 2019.

\bibitem[Gillman et~al.(2018{\natexlab{a}})Gillman, Wilson, Fuchi, Hartl,
  Pankonien, and Buskohl]{gillman2018design}
Andrew Gillman, Gregory Wilson, Kazuko Fuchi, Darren Hartl, Alexander
  Pankonien, and Philip Buskohl.
\newblock Design of soft origami mechanisms with targeted symmetries.
\newblock \emph{Actuators}, 8\penalty0 (1):\penalty0 3, 2018{\natexlab{a}}.

\bibitem[Gillman et~al.(2019)Gillman, Fuchi, and
  Buskohl]{gillman2019discovering}
Andrew~S Gillman, Kazuko Fuchi, and Philip~R Buskohl.
\newblock Discovering sequenced origami folding through nonlinear mechanics and
  topology optimization.
\newblock \emph{Journal of Mechanical Design}, 141\penalty0 (4):\penalty0
  041401, 2019.

\bibitem[Shende et~al.(2021)Shende, Gillman, Yoo, Buskohl, and
  Vemaganti]{shende2021bayesian}
Sourabh Shende, Andrew Gillman, David Yoo, Philip Buskohl, and Kumar Vemaganti.
\newblock Bayesian topology optimization for efficient design of origami
  folding structures.
\newblock \emph{Structural and Multidisciplinary Optimization}, 63:\penalty0
  1907--1926, 2021.

\bibitem[Lee et~al.(2024)Lee, Miyajima, and Tachi]{lee2024designing}
Munkyun Lee, Yuki Miyajima, and Tomohiro Tachi.
\newblock Designing and analyzing multistable mechanisms using quadrilateral
  boundary rigid origami.
\newblock \emph{Journal of Mechanisms and Robotics}, 16\penalty0 (1):\penalty0
  011008, 2024.

\bibitem[Feng et~al.(2020{\natexlab{a}})Feng, Dang, James, and
  Plucinsky]{feng2020designs}
Fan Feng, Xiangxin Dang, Richard~D. James, and Paul Plucinsky.
\newblock The designs and deformations of rigidly and flat-foldable origami.
\newblock \emph{Journal of the Mechanics and Physics of Solids}, page 104018,
  2020{\natexlab{a}}.
\newblock ISSN 0022-5096.

\bibitem[Feng et~al.(2020{\natexlab{b}})Feng, Plucinsky, and
  James]{feng2020helical}
Fan Feng, Paul Plucinsky, and Richard~D James.
\newblock Helical miura origami.
\newblock \emph{Physical Review E}, 101\penalty0 (3):\penalty0 033002,
  2020{\natexlab{b}}.

\bibitem[Dudte et~al.(2021)Dudte, Choi, and Mahadevan]{dudte2021additive}
Levi~H Dudte, Gary~PT Choi, and L~Mahadevan.
\newblock An additive algorithm for origami design.
\newblock \emph{Proceedings of the National Academy of Sciences}, 118\penalty0
  (21):\penalty0 e2019241118, 2021.

\bibitem[Lang and Howell(2018)]{lang2018rigidly}
Robert~J Lang and Larry Howell.
\newblock Rigidly foldable quadrilateral meshes from angle arrays.
\newblock \emph{Journal of Mechanisms and Robotics}, 10\penalty0 (2):\penalty0
  021004, 2018.

\bibitem[Dang et~al.(2022)Dang, Feng, Plucinsky, James, Duan, and
  Wang]{dang2022inverse}
Xiangxin Dang, Fan Feng, Paul Plucinsky, Richard~D James, Huiling Duan, and
  Jianxiang Wang.
\newblock Inverse design of deployable origami structures that approximate a
  general surface.
\newblock \emph{International Journal of Solids and Structures}, 234:\penalty0
  111224, 2022.

\bibitem[Dorn et~al.(2022)Dorn, Lang, and Pellegrino]{dorn2022kirigami}
Charles Dorn, Robert~J Lang, and Sergio Pellegrino.
\newblock Kirigami tiled surfaces with multiple configurations.
\newblock \emph{Proceedings of the Royal Society A}, 478\penalty0
  (2267):\penalty0 20220405, 2022.

\bibitem[Li and Pellegrino(2020)]{li2020theory}
Yang Li and Sergio Pellegrino.
\newblock A theory for the design of multi-stable morphing structures.
\newblock \emph{Journal of the Mechanics and Physics of Solids}, 136:\penalty0
  103772, 2020.

\bibitem[Dieleman et~al.(2020)Dieleman, Vasmel, Waitukaitis, and van
  Hecke]{dieleman2020jigsaw}
Peter Dieleman, Niek Vasmel, Scott Waitukaitis, and Martin van Hecke.
\newblock Jigsaw puzzle design of pluripotent origami.
\newblock \emph{Nature Physics}, 16\penalty0 (1):\penalty0 63--68, 2020.

\bibitem[Liu and James(2024)]{liu2024design}
Huan Liu and Richard~D James.
\newblock Design of origami structures with curved tiles between the creases.
\newblock \emph{Journal of the Mechanics and Physics of Solids}, page 105559,
  2024.

\bibitem[Dieleman(2018)]{dieleman2018origami}
Peter Dieleman.
\newblock \emph{Origami metamaterials: design, symmetries, and combinatorics}.
\newblock PhD thesis, Leiden University, 2018.

\bibitem[Zhou et~al.(2023)Zhou, Grasinger, Buskohl, and
  Bhattacharya]{zhou2023low}
Hao Zhou, Matthew Grasinger, Philip Buskohl, and Kaushik Bhattacharya.
\newblock Low energy fold paths in multistable origami structures.
\newblock \emph{International Journal of Solids and Structures}, 265:\penalty0
  112125, 2023.

\bibitem[Tachi(2009)]{tachi2009simulation}
Tomohiro Tachi.
\newblock Simulation of rigid origami.
\newblock \emph{Origami}, 4\penalty0 (08):\penalty0 175--187, 2009.

\bibitem[Li(2020)]{li2020motion}
Yang Li.
\newblock Motion paths finding for multi-degree-of-freedom mechanisms.
\newblock \emph{International Journal of Mechanical Sciences}, 185:\penalty0
  105709, 2020.

\bibitem[Fuchi et~al.(2016)Fuchi, Buskohl, Joo, Reich, and
  Vaia]{fuchi2016numerical}
K~Fuchi, PR~Buskohl, JJ~Joo, GW~Reich, and RA~Vaia.
\newblock Numerical analysis of origami structures through modified frame
  elements.
\newblock \emph{Origami6; Miura, K., Kawasaki, T., Tachi, T., Uehara, R., Lang,
  RJ, Wang-Iverson, P., Eds}, pages 385--395, 2016.

\bibitem[Ma and You(2014)]{ma2014energy}
Jiayao Ma and Zhong You.
\newblock Energy absorption of thin-walled square tubes with a prefolded
  origami pattern—part i: geometry and numerical simulation.
\newblock \emph{Journal of applied mechanics}, 81\penalty0 (1):\penalty0
  011003, 2014.

\bibitem[Zhang et~al.(2017)Zhang, Wommer, O’Rourke, Teitelman, Tang, Robison,
  Lin, and Yin]{zhang2017origami}
Qiuting Zhang, Jonathon Wommer, Connor O’Rourke, Joseph Teitelman, Yichao
  Tang, Joshua Robison, Gaojian Lin, and Jie Yin.
\newblock Origami and kirigami inspired self-folding for programming
  three-dimensional shape shifting of polymer sheets with light.
\newblock \emph{Extreme Mechanics Letters}, 11:\penalty0 111--120, 2017.

\bibitem[Gillman et~al.(2018{\natexlab{b}})Gillman, Fuchi, and
  Buskohl]{gillman2018truss}
Andrew Gillman, Kazuko Fuchi, and Phil~R Buskohl.
\newblock Truss-based nonlinear mechanical analysis for origami structures
  exhibiting bifurcation and limit point instabilities.
\newblock \emph{International Journal of Solids and Structures}, 147:\penalty0
  80--93, 2018{\natexlab{b}}.

\bibitem[Schenk and Guest(2011)]{schenk2011origami}
Mark Schenk and Simon~D Guest.
\newblock Origami folding: A structural engineering approach.
\newblock \emph{Origami}, 5:\penalty0 291--304, 2011.

\bibitem[Liu and Paulino(2017)]{liu2017nonlinear}
Ke~Liu and Glaucio~H Paulino.
\newblock Nonlinear mechanics of non-rigid origami: an efficient computational
  approach.
\newblock \emph{Proceedings of the Royal Society A: Mathematical, Physical and
  Engineering Sciences}, 473\penalty0 (2206):\penalty0 20170348, 2017.

\bibitem[Filipov et~al.(2017)Filipov, Liu, Tachi, Schenk, and
  Paulino]{filipov2017bar}
ET~Filipov, K~Liu, Tomohiro Tachi, Mark Schenk, and Glaucio~H Paulino.
\newblock Bar and hinge models for scalable analysis of origami.
\newblock \emph{International Journal of Solids and Structures}, 124:\penalty0
  26--45, 2017.

\bibitem[Lee-Trimble et~al.(2022)Lee-Trimble, Kang, Hayward, and
  Santangelo]{lee2022robust}
ME~Lee-Trimble, Ji-Hwan Kang, Ryan~C Hayward, and Christian~D Santangelo.
\newblock Robust folding of elastic origami.
\newblock \emph{Soft Matter}, 18\penalty0 (34):\penalty0 6384--6391, 2022.

\bibitem[Chen et~al.(2023)Chen, Xu, Lu, Liu, Feng, and Sareh]{chen2023multi}
Yao Chen, Ruizhi Xu, Chenhao Lu, Ke~Liu, Jian Feng, and Pooya Sareh.
\newblock Multi-stability of the hexagonal origami hypar based on group theory
  and symmetry breaking.
\newblock \emph{International Journal of Mechanical Sciences}, 247:\penalty0
  108196, 2023.

\bibitem[Chen and Santangelo(2018)]{chen2018branches}
Bryan Gin-ge Chen and Christian~D Santangelo.
\newblock Branches of triangulated origami near the unfolded state.
\newblock \emph{Physical Review X}, 8\penalty0 (1):\penalty0 011034, 2018.

\bibitem[McInerney et~al.(2020)McInerney, Chen, Theran, Santangelo, and
  Rocklin]{mcinerney2020hidden}
James McInerney, Bryan Gin-ge Chen, Louis Theran, Christian~D Santangelo, and
  D~Zeb Rocklin.
\newblock Hidden symmetries generate rigid folding mechanisms in periodic
  origami.
\newblock \emph{Proceedings of the National Academy of Sciences}, 117\penalty0
  (48):\penalty0 30252--30259, 2020.

\bibitem[belcastro and Hull(2002)]{hull2002modelling}
sarah belcastro and Thomas~C Hull.
\newblock Modelling the folding of paper into three dimensions using affine
  transformations.
\newblock \emph{Linear Algebra and its applications}, 348\penalty0
  (1-3):\penalty0 273--282, 2002.

\bibitem[Wu and You(2010)]{wu2010modelling}
Weina Wu and Zhong You.
\newblock Modelling rigid origami with quaternions and dual quaternions.
\newblock \emph{Proceedings of the Royal Society A: Mathematical, physical and
  engineering sciences}, 466\penalty0 (2119):\penalty0 2155--2174, 2010.

\bibitem[Zheng et~al.(2022)Zheng, Niloy, Celli, Tobasco, and
  Plucinsky]{zheng2022continuum}
Yue Zheng, Imtiar Niloy, Paolo Celli, Ian Tobasco, and Paul Plucinsky.
\newblock Continuum field theory for the deformations of planar kirigami.
\newblock \emph{Physical Review Letters}, 128\penalty0 (20):\penalty0 208003,
  2022.

\bibitem[Zheng et~al.(2023)Zheng, Niloy, Tobasco, Celli, and
  Plucinsky]{zheng2023modelling}
Yue Zheng, I~Niloy, I~Tobasco, P~Celli, and P~Plucinsky.
\newblock Modelling planar kirigami metamaterials as generalized elastic
  continua.
\newblock \emph{Proceedings of the Royal Society A}, 479\penalty0
  (2272):\penalty0 20220665, 2023.

\bibitem[Xu et~al.(2023)Xu, Tobasco, and Plucinsky]{xu2023derivation}
Hu~Xu, Ian Tobasco, and Paul Plucinsky.
\newblock Derivation of an effective plate theory for parallelogram origami
  from bar and hinge elasticity.
\newblock \emph{arXiv preprint arXiv:2311.10870}, 2023.

\bibitem[Huffman(1976)]{huffman1976curvature}
David~A. Huffman.
\newblock Curvature and creases: A primer on paper.
\newblock \emph{IEEE Transactions on computers}, \penalty0 (10):\penalty0
  1010--1019, 1976.

\bibitem[Hanna et~al.(2014)Hanna, Lund, Lang, Magleby, and
  Howell]{hanna2014waterbomb}
Brandon~H Hanna, Jason~M Lund, Robert~J Lang, Spencer~P Magleby, and Larry~L
  Howell.
\newblock Waterbomb base: a symmetric single-vertex bistable origami mechanism.
\newblock \emph{Smart Materials and Structures}, 23\penalty0 (9):\penalty0
  094009, 2014.

\bibitem[Hanna et~al.(2015)Hanna, Magleby, Lang, and Howell]{hanna2015force}
Brandon~H Hanna, Spencer~P Magleby, Robert~J Lang, and Larry~L Howell.
\newblock Force--deflection modeling for generalized origami waterbomb-base
  mechanisms.
\newblock \emph{Journal of Applied Mechanics}, 82\penalty0 (8), 2015.

\bibitem[Liu et~al.(2021)Liu, Plucinsky, Feng, and James]{liu2021origami}
Huan Liu, Paul Plucinsky, Fan Feng, and Richard~D James.
\newblock Origami and materials science.
\newblock \emph{Philosophical Transactions of the Royal Society A},
  379\penalty0 (2201):\penalty0 20200113, 2021.

\bibitem[James(2006)]{james2006objective}
Richard~D James.
\newblock Objective structures.
\newblock \emph{Journal of the Mechanics and Physics of Solids}, 54\penalty0
  (11):\penalty0 2354--2390, 2006.

\bibitem[Foschi et~al.(2022)Foschi, Hull, and Ku]{foschi2022explicit}
Riccardo Foschi, Thomas~C Hull, and Jason~S Ku.
\newblock Explicit kinematic equations for degree-4 rigid origami vertices,
  euclidean and non-euclidean.
\newblock \emph{Physical Review E}, 106\penalty0 (5):\penalty0 055001, 2022.

\bibitem[Farnham et~al.(2022)Farnham, Hull, and Rumbolt]{farnham2022rigid}
Johnna Farnham, Thomas~C Hull, and Aubrey Rumbolt.
\newblock Rigid folding equations of degree-6 origami vertices.
\newblock \emph{Proceedings of the Royal Society A}, 478\penalty0
  (2260):\penalty0 20220051, 2022.

\bibitem[Demaine et~al.(2011)Demaine, Demaine, Hart, Price, and
  Tachi]{demaine2011non}
Erik~D Demaine, Martin~L Demaine, Vi~Hart, Gregory~N Price, and Tomohiro Tachi.
\newblock (non) existence of pleated folds: how paper folds between creases.
\newblock \emph{Graphs and Combinatorics}, 27:\penalty0 377--397, 2011.

\bibitem[Liu et~al.(2019)Liu, Tachi, and Paulino]{liu2019invariant}
Ke~Liu, Tomohiro Tachi, and Glaucio~H Paulino.
\newblock Invariant and smooth limit of discrete geometry folded from bistable
  origami leading to multistable metasurfaces.
\newblock \emph{Nature communications}, 10\penalty0 (1):\penalty0 4238, 2019.

\bibitem[Inc.()]{Mathematica}
Wolfram~Research{,} Inc.
\newblock Mathematica, {V}ersion 13.3.
\newblock URL \url{https://www.wolfram.com/mathematica}.
\newblock Champaign, IL, 2023.

\bibitem[Yellowhorse et~al.(2018)Yellowhorse, Lang, Tolman, and
  Howell]{yellowhorse2018creating}
Alden Yellowhorse, Robert~J Lang, Kyler Tolman, and Larry~L Howell.
\newblock Creating linkage permutations to prevent self-intersection and enable
  deployable networks of thick-origami.
\newblock \emph{Scientific reports}, 8\penalty0 (1):\penalty0 12936, 2018.

\bibitem[Lang et~al.(2018)Lang, Tolman, Crampton, Magleby, and
  Howell]{lang2018review}
Robert~J Lang, Kyler~A Tolman, Erica~B Crampton, Spencer~P Magleby, and Larry~L
  Howell.
\newblock A review of thickness-accommodation techniques in origami-inspired
  engineering.
\newblock \emph{Applied Mechanics Reviews}, 70\penalty0 (1):\penalty0 010805,
  2018.

\bibitem[Chen et~al.(2015)Chen, Peng, and You]{chen2015origami}
Yan Chen, Rui Peng, and Zhong You.
\newblock Origami of thick panels.
\newblock \emph{Science}, 349\penalty0 (6246):\penalty0 396--400, 2015.

\bibitem[Zhu and Filipov(2024)]{zhu2024large}
Yi~Zhu and Evgueni~T Filipov.
\newblock Large-scale modular and uniformly thick origami-inspired adaptable
  and load-carrying structures.
\newblock \emph{Nature Communications}, 15\penalty0 (1):\penalty0 2353, 2024.

\bibitem[Zhu and Filipov(2019)]{zhu2019efficient}
Yi~Zhu and Evgueni~T Filipov.
\newblock An efficient numerical approach for simulating contact in origami
  assemblages.
\newblock \emph{Proceedings of the Royal Society A}, 475\penalty0
  (2230):\penalty0 20190366, 2019.

\bibitem[M{\"o}ller and Trumbore(2005)]{moller2005fast}
Tomas M{\"o}ller and Ben Trumbore.
\newblock Fast, minimum storage ray/triangle intersection.
\newblock In \emph{ACM SIGGRAPH 2005 Courses}, pages 7--es. 2005.

\end{thebibliography}

\end{document}


\title{Lagrangian approach to origami vertex analysis: Kinematics -- Supplementary Information}

\author{Matthew Grasinger}
\email{matthew.grasinger.1@us.af.mil}
\affiliation{Materials \& Manufacturing Directorate, Air Force Research Laboratory}

\author{Andrew Gillman}
\affiliation{Materials \& Manufacturing Directorate, Air Force Research Laboratory}

\author{Philip R. Buskohl}
\email{philip.buskohl.1@us.af.mil}
\affiliation{Materials \& Manufacturing Directorate, Air Force Research Laboratory}

\maketitle

\section{Preamble}

First recall the compatibility condition~\cite{hull2002modelling}
\begin{equation} \label{eq:compatibility}
  \F_{\NCreases}\left(\foldAngle_1, \dots, \foldAngle_{\NCreases}\right) = \prod_{i=1}^{\NCreases} \rotAbout{\CreaseVec_i}\left(\foldAngle_i\right) = \Iden,
\end{equation}
where the rotation transformations can be expressed using
\begin{equation} \label{eq:general-rotation}
  	\rotAbout{\unitVec}\left(\genAngle\right) = \exp\left(\genAngle \skwAbout{\unitVec}\right) = \left(1 - \cos \genAngle\right) \outProd{\unitVec}{\unitVec} + \cos \genAngle \Iden + \sin \genAngle \skwAbout{\unitVec}.
  \end{equation}
Although the primary focus of this work has been in obtaining and utilizing closed-form solutions, we show here that the Lagrangian approach also facilitates exploiting symmetry for reduced order computational folding.

\section{Reduced order computational folding} \label{app:computational}

Inspired by the computational folding approach first developed by Tachi~\cite{tachi2009simulation}, we show how the reduced order compatibility conditions can be linearized and approximated numerically.
To this end, consider a folded origami with flat state crease vectors, $\CreaseVec_i$, fold angles, $\foldAngle_i$, and folded configuration crease vectors, $\creaseVec_i$ for $i = 1, \dots, \NCreases$.
Let $\infFoldAngle_i$ denote a perturbation to the fold angle, $\foldAngle_i$, such that $\foldAngle_i \rightarrow \foldAngle_i + \infFoldAngle_i$.
A natural question is: ``given the folded state of the origami, what are the allowable perturbations, $\infFoldAngle_i, \: i = 1, \dots, \NCreases$?''.
Substituting the perturbed fold angles into \eqref{eq:compatibility}, and assuming $\prod_{i=1}^{\NCreases} \rotAbout{\CreaseVec_i}\left(\foldAngle_i\right) = \Iden$, one obtains
\begin{equation} \label{eq:compat-new}
	\prod_{i=1}^{\NCreases} \rotAbout{\CreaseVec_i}\left(\foldAngle_i + \infFoldAngle_i \right) = \prod_{i=1}^{\NCreases} \rotAbout{\creaseVec_i}\left(\infFoldAngle_i\right) = \Iden
\end{equation}
the significance of which is clear once one realizes that there is no requirement that the flat state be the choice of reference configuration.
One can, in principle, just as easily describe the deformation of the structure relative to a given folded configuration, in which case, \eqref{eq:compat-new} is the analogous compatibility condition.
However, \eqref{eq:compat-new} is of the same form as \eqref{eq:compatibility} and, hence, no easier to solve.
To make further progress, we will assume the fold angle perturbations are ``small enough''.
To this end, expanding \eqref{eq:general-rotation} to linear order about zero angle of rotation, we obtain the infinitesimal rotation tensor
\begin{equation} \label{eq:lin-rot}
	\rotAbout{\unitVec}\left(\infFoldAngle\right) = \Iden + \infFoldAngle \skwAbout{\unitVec} + \orderOf{\infFoldAngle^2}.
\end{equation}
Let $\smallParam = \max_i \left|\infFoldAngle_i\right|$.
Then, to linear order in $\smallParam$, the left hand side of \eqref{eq:compat-new} takes the form
\begin{equation}
	\prod_{i=1}^{\NCreases} \rotAbout{\creaseVec_i}\left(\infFoldAngle_i\right) = \prod_{i=1}^{\NCreases} \left(\Iden + \infFoldAngle_i \skwAbout{\creaseVec_i} + \orderOf{\infFoldAngle_i^2}\right) = \Iden + \sum_{i=1}^{\NCreases} \infFoldAngle_i \skwAbout{\creaseVec_i} + \orderOf{\smallParam^2}.
\end{equation}
Dropping higher order terms and setting this equal to the identity, we arrive at the linearized compatibility equation
\begin{equation} \label{eq:linear-Hull}
	\sum_{i=1}^{\NCreases} \infFoldAngle_i \skwAbout{\creaseVec_i} = \mathbf{0},
\end{equation}
which can readily be solved using numerical methods for finding matrix nullspaces.
Computational discovery of a folding trajectory can be accomplished iteratively by, at each step, solving for possible perturbations to the fold angles, updating the fold angles via numerical integration and crease vectors via the deformation map, and then repeating the process about each new folded configuration~\cite{tachi2009simulation}.
This process is analogous to the \emph{updated Lagrangian} formulation in continuum solid mechanics.

\paragraph*{Computational folding while imposing a reflection symmetry.}

Given the overview of an algorithm for computational folding, let us now return our attention to the reduced order formulation for reflection symmetries (section IV of the main text).
Here we consider the folded origami, $\cBody$, as our new choice of reference configuration, where the $\creaseVec_i$ satisfy have a reflection symmetry about the plane orthogonal to $\etwo$.
Let the deformation map be given by
\begin{equation}
	\deffMap\left(\xC\right) = \begin{cases}
		\rotate{\creaseVec_1}{\frac{\infFoldAngle_1}{2}} \xC, & \xC \in \facet{1} \\
		\rotate{\creaseVec_1}{\frac{\infFoldAngle_1}{2}} \prod_{i=2}^{j} \rotate{\creaseVec_i}{\infFoldAngle_i} \xC, & \xC \in \facet{j}, j \neq 1 \\
	\end{cases}.
\end{equation}
Using arguments similar to those given in the main text, we arrive at
\begin{subequations}
\begin{align}
	\label{eq:inf-compatibility}
	\F'_{M} &\coloneqq \rotate{\creaseVec_1}{\frac{\infFoldAngle_1}{2}} \prod_{i=2}^{M} \rotate{\creaseVec_i}{\infFoldAngle_i}, \\
	\F'_{\NCreases} &= \rotate{\creaseVec_1}{-\frac{\infFoldAngle_1}{2}}, \\
	\label{eq:inf-reflection-condition-1}
	\left(\F'_{\NCreases / 2 - 1} \creaseVec_{\NCreases / 2}\right) \cdot \etwo &= 0,
\end{align}
\end{subequations}
as the definitions and conditions for $\infFoldAngle_i, i = 1, \dots \NCreases$ to describe a compatible configuration with a reflection symmetry about the plane orthogonal to $\etwo$.
Substituting in the rotation expansion, \eqref{eq:lin-rot}, and dropping higher order terms, we obtain the linear, symmetry reduced order condition
\begin{equation} \label{eq:inf-reflection-condition}
	\begin{split}
		\left(\left(\Iden + \frac{\infFoldAngle_1}{2} \skwAbout{\creaseVec_1}\right) \left(\Iden + \infFoldAngle_2 \skwAbout{\creaseVec_2}\right) \dots \left(\Iden + \infFoldAngle_{\NCreases/2 - 1} \skwAbout{\NCreases/2 - 1}\right) \creaseVec_{\NCreases / 2}\right) \cdot \etwo + \orderOf{\smallParam^2} &= 0, \\
		\left(\left(\Iden + \frac{\infFoldAngle_1}{2} \skwAbout{\creaseVec_1} + \sum_{i=2}^{\NCreases / 2 - 1} \infFoldAngle_i \skwAbout{\creaseVec_i}\right) \creaseVec_{\NCreases / 2}\right) \cdot \etwo &= 0,
	\end{split}
\end{equation}
A numerical solution can be obtained by iteratively \begin{enumerate} 
	\item choosing $\infFoldAngle_1, ..., \infFoldAngle_{\NCreases / 2 - 1}$ such that \eqref{eq:inf-reflection-condition} is satisfied and $\sqrt{\infFoldAngle_1^2 + \dots \infFoldAngle_{\NCreases / 2 - 1}^2} = 1$, \item letting $\infFoldAngle_{\NCreases / 2 + 1} = \infFoldAngle_{\NCreases / 2 - 1}, \dots, \infFoldAngle_{\NCreases} = \infFoldAngle_2$, \item performing an integration step (e.g. Euler integration: $\foldAngle_i \rightarrow \foldAngle_i + \stepSize \infFoldAngle_i$ where $\stepSize$ is the size of the Euler step), and \item updating the crease unit vectors (i.e. $\creaseVec_i \rightarrow \F_{i}\left(\foldAngle_1, \dots, \foldAngle_\NCreases\right) \CreaseVec_i$). 
\end{enumerate}
While forward Euler integration is simple to implement, we remark that, based on numerical experiments, $4$th order Runge-Kutta is very well suited for this approach, both in terms of efficiency and stability.

\paragraph*{Computational folding while imposing reflections and rotations.}
Consider next the computational folding of vertices with reflection and rotational symmetries.
Let a folded state be given such that $\genAngle = 2 \pi / 2 h, h \in \Naturals$, $\group = \dihedral{2 h}$, $\creaseVec_k = \rotate{\ethree}{\genAngle / 2} \creaseVec_1$ (i.e. $\refl{\creaseVec_k \times \ethree} \in \group$), and $\facet{c} = \bigcup_{i=1}^{k-1} \facet{i}$, where $\facet{c} \subset \cBody$ is the unit cell to be perturbed such that $\cBody = \orbit{\facet{c}}{\group}$.
The perturbation in the elevation angle is denoted by $\infElevAngle$.
Then the deformation map with respect to the folded configuration of interest is given by \begin{equation}
	\deffMap\left(\xC\right) = \begin{cases}
		\rotate{\eone}{\infElevAngle} \rotate{\creaseVec_1}{\frac{\infFoldAngle_1}{2}} \xC, & \xC \in \facet{1} \\
		\rotate{\eone}{\infElevAngle} \rotate{\creaseVec_1}{\frac{\infFoldAngle_1}{2}} \prod_{i=2}^{j} \rotate{\creaseVec_i}{\infFoldAngle_i} \xC, & \xC \in \facet{j}, j \neq 1 \\
	\end{cases}.
\end{equation}
The analog of the rotation-reflection compatibility condition (equation 5.15 of the main text) where the ``reference'' configuration is a folded state is given by:
\begin{equation}
	\frac{\left(\F'_{k-1}\left(\infElevAngle, \infFoldAngle_1, \dots, \infFoldAngle_{k-1}\right) \creaseVec_{k}\right) \cdot \etwo}{\left(\F'_{k-1}\left(\infElevAngle, \infFoldAngle_1, \dots, \infFoldAngle_{k-1}\right) \creaseVec_k\right) \cdot \eone} = \frac{\creaseVec_k \cdot \etwo}{\creaseVec_k \cdot \eone}.
\end{equation}
Substituting in \eqref{eq:lin-rot} and rearranging, one obtains
\begin{equation}
	\begin{split}
	&\left[\left(\left(\Iden + \infElevAngle \skwAbout{\eone} + \sum_{i=1}^{k-1} \infFoldAngle_i\skwAbout{\creaseVec_i}\right) \creaseVec_{k}\right) \cdot \etwo\right]\left(\creaseVec_k \cdot \eone\right) \\
	& \quad - \left[\left(\left(\Iden + \infElevAngle \skwAbout{\eone} + \sum_{i=1}^{k-1} \infFoldAngle_i\skwAbout{\creaseVec_i}\right) \creaseVec_{k}\right) \cdot \eone\right]\left(\creaseVec_k \cdot \etwo\right) = 0,
	\end{split}
\end{equation}
which, again, is linear in the unknowns.
Thus, it can iteratively be solved, numerical integration can be used to update the fold angles, and the deformation map can be used to update the crease vectors.

\section{Summary List of Animations} \label{app:animations}

\begin{itemize}
            \item \textbf{Animation \#1:} \texttt{6-fold vertex\_alpha1-pi\_6.avi} and \texttt{.gif}\
             
            Folding animation of a $6$-fold vertex with $\sectAngle_1 = \pi / 6$ along the trajectory highlighted by the location in the kinematic domain denoted by the moving star (see figure 3).
            \item \textbf{Animation \#2:} \texttt{6-fold vertex\_alpha1-pi\_4.avi} and \texttt{.gif}\
            
            Folding animations of a $6$-fold vertex with $\sectAngle_1 = \pi / 4$ folding along $\left(\foldAngle_2 = 0 \rightarrow -\pi, \foldAngle_3 = \pi\right)$ and then $\left(\foldAngle_2 = -\pi, \foldAngle_3 = \pi \rightarrow 0\right)$ (see figure 4b).
            \item \textbf{Animation \#3:} \texttt{6-fold vertex\_alpha1-5pi\_12.avi} and \texttt{.gif}\

            Folding animations of a $6$-fold vertex with $\sectAngle_1 = 5 \pi / 12$ folding along $\left(\foldAngle_2 = 0 \rightarrow -\pi, \foldAngle_3 = \pi\right)$ and then $\left(\foldAngle_2 = -\pi, \foldAngle_3 = \pi \rightarrow 0\right)$. Animations \#3 and \#4 are included to highlight why the admissible regions of configuration space are different between these two cases (see figure 4b).

            \item \textbf{Animation \#4:} \texttt{6-fold vertex\_alpha1-11pi\_24.avi} and \texttt{.gif}\
            
            Animation of the hinge-like behavior observed in the $6$-fold vertex when $\sectAngle_1$ approaches $\pi / 2$, as shown in this $6$-fold, $\sectAngle_1 = 11 \pi / 24$ vertex (see figure 4a and 4b).
            \item \textbf{Animation \#5:} \texttt{8-fold vertex\_alpha1-pi\_6.avi} and \texttt{.gif}\
            
            Folding animation of an $8$-fold vertex with $\sectAngle_1 = \pi / 6$ and $\dihedral{2}$ symmetry. To highlight the different regions in configuration space, the vertex is folded into admissible regions, regions where no solution exists, and regions where contact occurs (see figure 9).
        \end{itemize}

\bibliographystyle{alpha}
\bibliography{master}